\documentclass[apj]{emulateapj}
\newcommand{\mpch}{\>h^{-1}{\rm {Mpc}}}

\newcommand{\FIred}{\rm F1_{red}}
\newcommand{\FIIred}{\rm F2_{red}}
\newcommand{\FIIIred}{\rm F3_{red}}

\newcommand{\FVred}{\rm F5_{red}}
\newcommand{\FVIred}{\rm F6_{red}}
\newcommand{\FIblue}{\rm F1_{blue}}

\newcommand{\FVIblue}{\rm F6_{blue}}

\begin{document}

\shorttitle{The cross-correlation between galaxies}
\shortauthors{Wang et al.}

\title{The cross-correlation between galaxies
of different luminosities and colors}

\author{Yu Wang\altaffilmark{1,3}, Xiaohu Yang\altaffilmark{1,3},
        H.J. Mo\altaffilmark{2}, Frank C. van den Bosch\altaffilmark{4}}

\altaffiltext{1}{Shanghai Astronomical Observatory,
      the Partner Group of MPA, Nandan Road 80, Shanghai 200030, China}
\altaffiltext{2}{Department of Astronomy, University of Massachusetts,
      Amherst MA 01003-9305}
\altaffiltext{3}{Joint Institute for Galaxy and Cosmology (JOINGC) of
      Shanghai Astronomical Observatory and University of Science and
      Technology of China}
\altaffiltext{4}{Max-Planck Institute for Astronomy,
       D-69117 Heidelberg, Germany}

\begin{abstract}
  We  study  the   cross-correlation  between  galaxies  of  different
  luminosities and colors, using  a sample of 284489 galaxies selected
  from the  Sloan Digital  Sky Survey, Data  Release 4.   Galaxies are
  divided into  6 samples according  to luminosity, and each  of these
  samples is further divided into red and blue subsamples according to
  their $(g-r)$  colors.  Projected auto-correlation  is estimated for
  each  subsample, and  projected cross-correlation  is  estimated for
  each pair of subsamples.  At projected separations $r_p \ga 1\mpch$,
  all  correlation  functions   are  roughly  parallel,  although  the
  correlation amplitude depends  systematically on both luminosity and
  color. On $r_p\la 1\mpch$, the auto- and cross-correlation functions
  of  red   galaxies  are  significantly  enhanced   relative  to  the
  corresponding   power  laws   obtained  on   larger   scales.   Such
  enhancement is absent for blue galaxies and in the cross-correlation
  between red and blue galaxies.  We estimate the relative bias factor
  on   scales   $r\ga   1\mpch$   for   each   subsample   using   its
  auto-correlation function as well as its cross-correlation functions
  with  other  subsamples. The  relative  bias  factors obtained  from
  different   methods   are    similar.    For   blue   galaxies   the
  luminosity-dependence of the relative bias is strong over the entire
  luminosity  range  probed  ($-23.0<M_r\le  -18.0$),  while  for  red
  galaxies  the dependence  is  weaker and  becomes insignificant  for
  luminosities below $L^*$.  In order to examine whether a significant
  stochastic/nonlinear component exists in the bias relation, we study
  the   ratio  ${\cal   R}_{ij}\equiv   W_{ii}W_{jj}/W_{ij}^2$,  where
  $W_{ij}$ is  the projected  correlation between subsample  `$i$' and
  `$j$'. We find that the values of ${\cal R}_{ij}$ are all consistent
  with  1   for  all-all,  red-red  and   blue-blue  samples,  however
  significantly larger  than 1 for  red-blue samples. For faint  red -
  faint  blue samples the  values of  ${\cal R}_{ij}$  are as  high as
  $\sim  3$  on  small  scales  ($r_p\la 1\mpch$)  and  decrease  with
  increasing   $r_p$.  These  results   suggest  that   a  significant
  stochastic/nonlinear  component exists  in the  relationship between
  red and blue galaxies, particularly on small scales.
\end{abstract}

\keywords{large-scale structure of the universe - galaxies:
halos - methods: statistical}

\section{Introduction}

In the  current scenario of structure formation,  galaxies are assumed
to  form in  the cosmic  density field  through a  series  of physical
processes, such as non-linear  gravitational collapse, gas cooling and
star formation. In this scenario the distribution of galaxies in space
is expected to trace the  underlying density field to some degree, but
the relationship  is not expected to  be perfect, because  many of the
processes involved in galaxy formation can complicate the relationship
between galaxies and  the dark matter density field.   Thus, the study
of  galaxy  clustering   in  space  is  an  important   step  both  in
understanding  the   mass  distribution   in  the  universe,   and  in
understanding  how galaxies  form in  the cosmic  density  field.  The
statistical  tool commonly  adopted to  quantify galaxy  clustering in
space is the correlation  functions of galaxies (Peebles 1980). During
the  last  two decades,  many  authors  have  estimated the  two-point
correlation functions of galaxies  using various redshift catalogs, to
study  how galaxy clustering  depends on  the properties  of galaxies,
such as luminosity (e.g.  B\"orner, Mo \& Zhou 1989; Park et al. 1994;
Loveday et al.  1995; Guzzo et al. 1997; Benoist  et al. 1996; Norberg
et al.  2001; Zehavi et al. 2002; Zehavi et al. 2005; Li et al. 2006),
color (e.g.  Willmer et  al. 1998; Brown  et al.  2000; Zehavi et  al.
2002; Zehavi et  al. 2005; Li et al. 2006),  spectral type (Norberg et
al.  2002; Budavari et al.  2003; Madgwick et al. 2003), morphological
type (e.g.  Jing, Mo \& B\"orner  1991; Guzzo et al.  1997; Willmer et
al.  1998; Zehavi  et  al. 2002;  Goto  et al.   2003), stellar  mass,
stellar surface mass density and  concentration (Li et al.  2006). All
these  analyses show that  galaxies of  different properties  may have
different distributions in space,  indicating that galaxies are biased
tracers of the underlying density field and that one has to be careful
when  using  the  observed  galaxy  distribution  to  infer  the  mass
distribution in the universe.

The observed  correlation functions of galaxies  can provide important
constraint on the relationship between galaxies and dark matter halos,
as  is   demonstrated  clearly  in  the   recent  halo-occupation  and
conditional-luminosity-function  models (Jing,  Mo  \& B\"orner  1998;
Seljak 2000; Peacock  \& Smith 2000; Berlind \&  Weinberg 2002; Cooray
\& Sheth 2002; Yang, Mo \& van  den Bosch 2003; van den Bosch, Yang \&
Mo 2003). With accurate  measurements of the clustering properties for
galaxies of different properties, it is now possible to study in great
detail how different galaxies  occupy different halos and how galaxies
trace the cosmic density field.

The correlation  analyses carried out so  far are mostly  based on the
auto-correlation  function  of galaxies.  Although  such analyses  can
provide important  information about how  a population of  galaxies is
distributed  in  space,  they  do  not  tell  us  anything  about  the
relationship  between galaxies of  different properties.  For example,
two populations  of galaxies may  each be strongly clustered  in space
and yet be spatially  segregated so that the cross-correlation between
them is  weak. In reality, all  populations of galaxies  may be biased
tracers of the  underlying mass distribution, and so  they must all be
positively  correlated  in  space.   However,  the  amplitude  of  the
cross-correlation between any two  populations of galaxies depends not
only on  the bias factors  of these two  populations, but also  on how
well  each  of  them  trace  the  mass  density  field,  i.e.  on  the
stochasticity in the bias relation.   Thus, it is important to measure
the   cross-correlation  functions   between  galaxies   of  different
properties, so  as to quantify  the relationships between  the spatial
distributions of different galaxies.

In  this paper,  we  use  galaxy samples  constructed  from the  Sloan
Digital  Sky Survey  Data Release  4  (SDSS-DR4) to  measure both  the
auto-correlation functions for  galaxies of different luminosities and
colors and the cross-correlation functions between different galaxies.
We compare  the shapes of  the various correlation functions  to check
the validity of the assumption  of linear bias. For each population of
galaxies  (i.e. for  galaxies within  given ranges  of  luminosity and
color),   we   estimate   its   relative  bias   factors   using   its
auto-correlation function as  well all its cross-correlation functions
with other populations of galaxies. These bias factors are averaged to
give  a mean  bias  factor  for each  population,  and compared  among
themselves to constrain possible stochastic/nonlinear component in the
bias relation (Mo  \& White 1996; Dekel \& Lahav  1999). This paper is
organized  as  follows. In  section  \ref{sec_data},  we describe  the
galaxy samples to  be used and how random  samples are constructed. In
section \ref{sec_method},  we describe  our method for  estimating the
correlation   functions.  Our  results   about  the   luminosity-  and
color-dependence  of  the   correlation  functions  are  presented  in
Sections  \ref{sec:result_L} and \ref{sec:result_c},  respectively. In
Section  \ref{sec:discussion}  we  discuss  the  implications  of  our
results for  the possible  existence of stochastic/non-linear  bias in
galaxy distribution.   Finally, we  summarize our findings  in Section
\ref{sec:summary}.

Throughout this  paper, galaxy  distances are obtained  from redshifts
assuming a  cosmology with $\Omega_{\rm  m, 0}=0.3$, $\Omega_{\Lambda,
  0}=0.7$.  Distances are  quoted in  units of  $\mpch$,  and absolute
magnitudes are quoted in terms of $M-5\log h$, i.e. with $h$ set to be
1, where $h=H_0/(100 {\rm km s^{-1}Mpc^{-1}})$.

\section{The data}
\label{sec_data}

\begin{figure*}
  \plotone{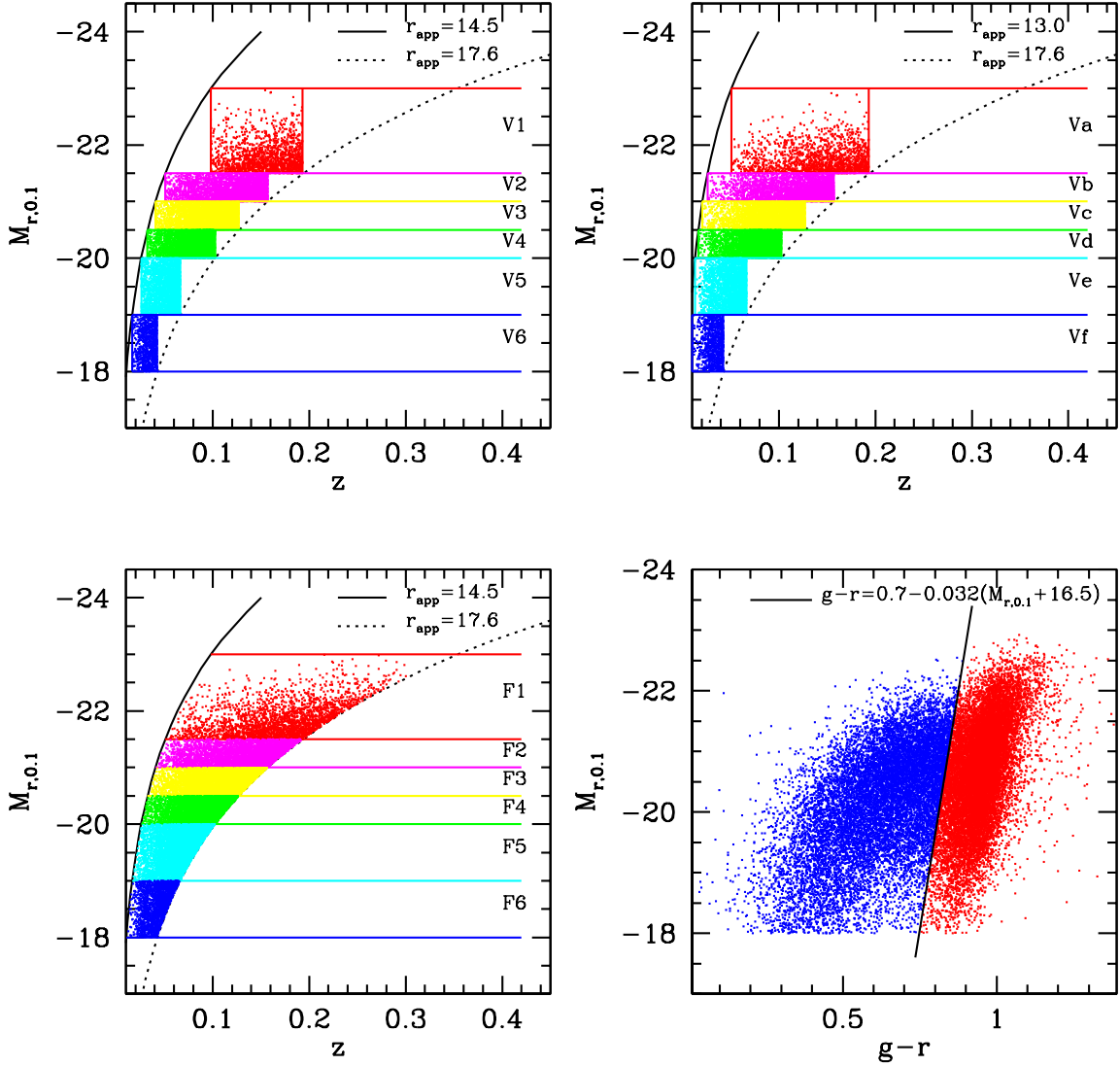}  \caption{The  magnitude-redshift distributions  of
    10\%  galaxies in  the volume-limited  samples V1--V6  (upper left
    panel), Va --  Vf (upper right panel) and  flux-limited samples F1
    -- F6  (lower  left  panel).   The  lower right  panel  shows  the
    color-magnitude  distribution   of  $10\%$  of   the  main  sample
    galaxies.   The   solid   line  $g-r=0.7-0.032   (M_{r,0.1}+16.5)$
    separates the  galaxies into red and blue  populations. See Tables
    ~\ref{tab1} and ~\ref{tab2} for more infomation. }
\label{fig:mgr}
\end{figure*}

\subsection{The SDSS Data Release 4}
\label{sec:data_sdss}

\begin{deluxetable*}{lccccc}
  \tablecaption{Volume-limited samples\label{tab1}} \tablewidth{0pt}
  \tablehead{ &&\multicolumn{3}{c}{Number of Galaxies} \\
    \cline{3-5} Sample & $M_{r,0.1}$ & NGCE & NGCO & SGC & z} \startdata
  V1/Va&(-23.0,~-21.5] &4670/5365   &8935/10030   &2704/3050 &(0.098,~0.193)/(0.051,~0.193)\\
  V2/Vb&(-21.5,~-21.0] &9288/9485   &16978/17442 &4777/4919 &(0.051,~0.157)/(0.026,~0.157)\\
  V3/Vc&(-21.0,~-20.5] &11833/12110 &19987/20542 &5676/5837 &(0.041,~0.127)/(0.021,~0.127)\\
  V4/Vd&(-20.5,~-20.0] &10842/11158 &17705/18348 &5146/5317 &(0.033,~0.103)/(0.017,~0.103)\\
  V5/Ve&(-20.0,~-19.0] &6493/6974   &15462/16073 &4317/4571 &(0.026,~0.066)/(0.013,~0.066)\\
  V6/Vf&(-19.0,~-18.0] &2620/2690   &5321/5485   &1534/1590 &(0.017,~0.043)/(0.010,~0.043)

  \enddata \tablecomments{Column 1 indicates the sample ID. Column 2 gives the
    absolute-magnitude range covered by each sample.  Columns 3, 4 and 5 list
    the galaxy number in each sample in the three regions (NGCE, NGCO, SGC) of
    the SDSS. Column 6 lists the redshift range of each sample. Each sample is
    a volume-limited in the redshift range it covers.  Samples V1 -- V6 are
    for galaxies with $r$-band apparent magnitudes in the range from 14.5 to
    17.6. Samples Va -- Vf are constructed from all galaxies with $r$-band
    apparent magnitudes in the range from 13.0 to 17.6. The reason for using
    these two sets of volume-limited samples is given in the main text.}
\end{deluxetable*}

\begin{deluxetable*}{lccccc}
  \tablecaption{Flux-limited samples \label{tab2}} \tablewidth{0pt}
  \tablehead{ &&\multicolumn{3}{c}{Number of Galaxies} \\
    \cline{3-5} Sample & $M$ & NGCE~(Red/Blue) & NGCO~(Red/Blue) &
    SGC~(Red/Blue) & $z$} \startdata
  F1~(${\rm F1_{red}}$/${\rm F1_{blue}}$) &(-23.0,~-21.5] &8817~(7203/1614)   &17012~(13745/3267)  &5108~(4147/961)&(0.051,~0.300)\\
  F2~(${\rm F2_{red}}$/${\rm F2_{blue}}$) &(-21.5,~-21.0] &14189~(9173/5016)  &26598~(17150/9448) &7556~(4938/2618)&(0.041,~0.193)\\
  F3~(${\rm F3_{red}}$/${\rm F3_{blue}}$) &(-21.0,~-20.5] &19025~(10968/8057) &33459~(19174/14285) &9498~(5434/4064)&(0.033,~0.157)\\
  F4~(${\rm F4_{red}}$/${\rm F4_{blue}}$) &(-20.5,~-20.0] &17214~(8979/8235)  &29037~(15186/13851) &8329~(4271/4058)&(0.026,~0.127)\\
  F5~(${\rm F5_{red}}$/${\rm F5_{blue}}$) &(-20.0,~-19.0] &20252~(8949/11303) &36386~(15810/20576) &10162~(4369/5793)&(0.017,~0.103)\\
  F6~(${\rm F6_{red}}$/${\rm F6_{blue}}$) &(-19.0,~-18.0] &5597~(1409/4188) &12687~(3570/9117)  &3563~(1041/2522)&(0.010,~0.066) \\
  F*         & (-20.7,~-20.2] &  18444 &  31714 & 8975 &

  \enddata   \tablecomments{Similar   to   Table~\ref{tab1}  but   for
    flux-limited samples.  Here again galaxies  with $r$-band apparent
    magnitudes in  the range from  14.5 to 17.6 are  selected. Numbers
    are listed separately for all,  red and blue galaxies. ${\rm F^*}$
    is a sample of galaxies with luminosities within a $0.5$ magnitude
    bin around $L^*$.}
\end{deluxetable*}

\begin{figure*}
  \plotone{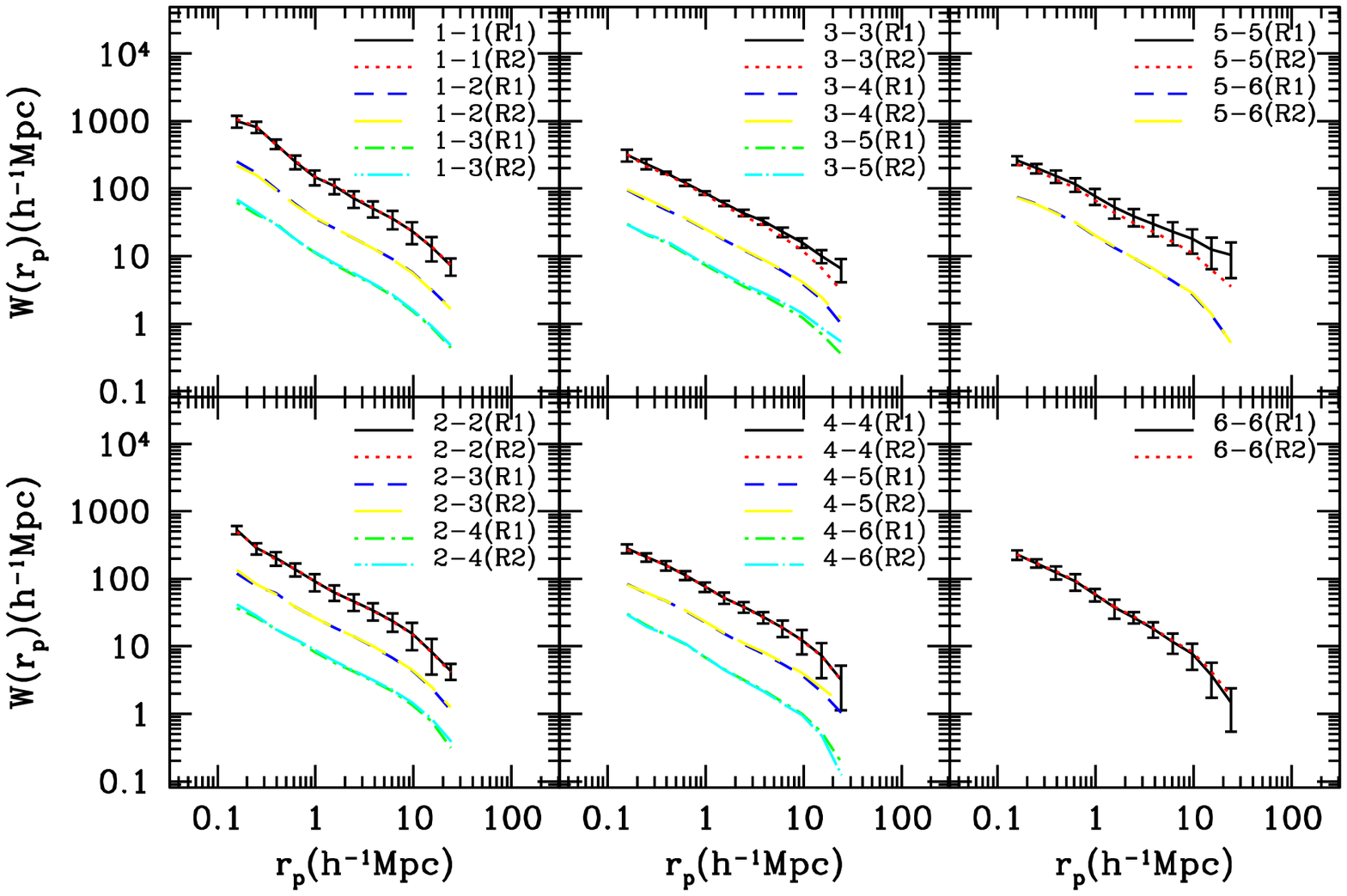} \caption{The  projected 2PCFs for  the flux-limited
    samples.   Here  we compare  the  results  obtained  by using  two
    different random samples, R1 and R2  (see the main text for how R1
    and R2  are constructed).  The  error-bars in this  plot represent
    the 1-$\sigma$ scatter among  the three SDSS regions. For clarity,
    we  only plot  error-bars for  the auto-correlation  functions but
    note that the error-bars  for the cross-correlation functions have
    similar  sizes.  The results  for the  cross-correlation functions
    are shifted  downward by  $0.5$ and  $1$ dex for  the cases  Fi --
    F(i+1)  and Fi-F(i+2) (i=1,2,3,4,5),  respectively.  Note  that in
    most of  the cases, these two  types of random  samples yield very
    similar  results.  However, discrepancy  exists  in  a few  cases,
    especially for the auto-correlation  function of F5.  As discussed
    in  the  main text,  this  discrepancy  is  likely caused  by  the
    supercluter at redshift $z\sim 0.08$.}
\label{auto}
\end{figure*}

In this paper we use galaxy samples constructed from the New York University
Value-Added Galaxy Catalog (NYU-VAGC, see Blanton et al. 2005). The version of
the NYU-VAGC used here is based on the SDSS Data Release 4 (Adelman-McCarthy
et al.  2006 ) which consists of three regions: two in the Northern Galactic
Cap (NGC) and one in the Southern Galactic Cap (SGC). We refer the two NGC
regions separately as NGCE (which is along the celestial equator) and NGCO
(which is off the equator). The SGC contains three stripes: one along the
celestial equator, one to the north and one to the south. In this paper, we
measure the correlation functions separately for NGCE, NGCO and SGC, and
obtain the variance from these samples. While the total correlation functions
of the whole samples are obtained by combining the pair counts in all these
three regions. Note that these three regions have quite large angular
separations which ensure us to obtain a good estimation of the cosmic
variance, except for those faintest galaxies at very low redshift where the
variance may be somewhat underestimated.  Our main sample contains 284489
galaxies with extinction-corrected Petrosian magnitudes in the range $14.5 < r
<17.6$, redshifts in the range $0.01<z<0.3$ and absolute magnitudes in the
range $-23.0< M_{r,0.1} <-18.0$.  The $r$-band absolute magnitudes are
corrected to redshift $z=0.1$ using the $K$-correction code of Blanton et
al.(2003a) and the luminosity-evolution model ($E$-correction) of Blanton et
al.(2003b), given by
\begin{equation}
M_{r,0.1} = M_{r,z} + 1.62(z-0.1)\,.
\end{equation}
In what  follows, all absolute magnitudes and  luminosities are quoted
with  such K-  and  E-corrections.  In  the apparent-magnitude  limits
adopted here,  $14.5 <  r <17.6$,  the bright end  is chosen  to avoid
incompleteness due to  the large angular sizes of  galaxies, while the
faint end  is chosen  to match  the magnitude limit  of the  SDSS main
galaxy   sample.  In   our  analysis,   we  will   also   use  another
apparent-magnitude  criterion,  $13.0  <   r  <17.6$  to  construct  a
supplemental sample.  This sample has in total  287728 galaxies, about
3239  more  than  the  main  sample described  above.  Most  of  these
additional  galaxies  are   intrinsically  bright  galaxies  with  low
redshifts, and the use of these galaxies can significantly improve the
signal  of the  cross-correlation function  between the  brightest and
faintest samples by increasing  their overlapping volume between them.
As we  will see  below, since we  can normalize  the cross-correlation
function  with the random  sample constructed  according to  the faint
sample that  is complete, and  since the incompleteness  introduced by
including the brightest galaxies is not expected to be correlated with
the   real  structure,   the  incompleteness   should   not  introduce
significant bias in our results.

From  our main  sample, we  first construct  6  volume-limited samples
(Samples V1 -- V6) in six bins of absolute-magnitude ($M_{r,0.1}$): V1
(-23.0,-21.5], V2  (-21.5,-21.0], V3 (-21.0,-20.5],  V4 (-20.5,-20.0],
V5 (-20.0,-19.0],  V6 (-19.0,-18.0], with  detailed selection criteria
given in  Table \ref{tab1}.   Obviously, the redshift  overlap between
the brightest and  faintest samples is small (and  sometimes none). In
order to enlarge the redshift  overlap, which is necessary in order to
measure  the   cross  correlations  reliably,  we   also  construct  6
volume-limited samples,  Va -- Vf,  from our supplemental  sample.  In
the upper panels of  Fig.~\ref{fig:mgr} we show the magnitude-redshift
distributions  of 10\%  of  all these  volume-limited samples:  V1--V6
(upper left panel) and Va -- Vf (upper right panel).

In  addition to the  volume-limited samples  described above,  we also
construct 6 flux-limited  samples, F1 -- F6. As for  samples V1 -- V6,
these are  defined by the apparent  magnitude limits $14.5  < r <17.6$
and by a bin in absolute  magnitude. However, no limits are imposed on
the   redshifts,   as  illustrated   in   the   lower-left  panel   of
Fig.~\ref{fig:mgr}.   The  details  of  these  samples  are  given  in
Table~\ref{tab2},  where the  last  column indicates  the minimum  and
maximum redshifts of  the galaxies in each sample  (which simply arise
because  of the apparent  magnitude limits).   The advantage  of using
flux-limited samples is that we have more galaxies in each sample, but
the drawback  is that  the samples are  not homogeneous in  the radial
direction, and so the two samples to be cross-correlated do not occupy
exactly the same  volume. As we will show below,  the results based on
the  flux-limited samples  are  very  similar to  those  based on  the
volume-limited samples.   Because of this,  we will focus on  the flux
samples to take advantage of the larger sample sizes.  For each of the
flux-limited  sample,  we  further  divide  galaxies  into  two  color
subsamples according to their $g-r$ colors. The dividing line is given
by
\begin{equation}\label{color_criterion}
g-r=0.7-0.032 (M_{r,0.1}+16.5)\,,
\end{equation}
where $g$  and $r$ are the  $g$-band and $r$-band  magnitudes that are
$K$-corrected (Blanton et al. 2003a) and $E$-corrected (Blanton et al.
2003b) to  redshift $z=0.1$. Galaxies  with $g-r$ color  above (below)
this  line  are called  red  (blue)  galaxies,  and the  corresponding
samples  are denoted  by $\FIred$  -- $\FVIred$  (red  subsamples) and
$\FIblue$ -- $\FVIblue$ (blue  subsamples). The selection criteria for
these subsample  are listed in  Table \ref{tab2}.  In the  lower right
panel of Fig.~\ref{fig:mgr} we show the color-magnitude relation for a
random set  of $10\%$  of all  galaxies in our  main sample,  with the
solid    line    indicating   the    color    separation   given    by
Eq.~(\ref{color_criterion}). Note  that this color  separation is very
similar to  what has been used  in previous studies  (e.g.  Blanton et
al. 2003c; Baldry  et al.  2004; Bell et al.  2004;  Hogg et al. 2004;
Weinmann et al.  2006).

As  comparison,  we  also  construct  a reference  sample,  F*,  which
includes all  galaxies within a $0.5$-magnitude bin  around $L^*$. The
selection  criteria for this  sample are  listed in  the last  line of
Table \ref{tab2}.

\subsection{The random samples}\label{sec:data_rand}

In order to measure the  two-point correlation functions, one needs
to construct random samples to normalize the galaxy-pair counts.
Since we will study the clustering properties  separately for all,
red and blue galaxies, we  need to generate  random samples
separately for  each of them. In general,  if we know the luminosity
function of galaxies, we can first construct  a random sample with
the  luminosity function and then apply  the observational magnitude
limits. Unfortunately,  we do not  have the  luminosity functions
for the  red  and blue  galaxies defined according to the color
criterion described above.  Because of this, here  we adopt a method
proposed  in Li et al.   (2006), where a random sample is generated
by assigning each galaxy in the real sample a random position in the
sky while keeping all other properties (i.e., redshift,  color,
magnitude,  etc.)   unchanged.  In  what follows we refer to the
random samples thus constructed as random samples of type R1. Note
that these random  samples have, by construction, exactly the same
redshift   distribution,  $n(z)$,   as  the original   sample.
Consequently, it is  not exactly `random' in that it can still
reveal structures in $n(z)$. Therefore, this  method is only
expected to work well for wide-angle  surveys in which the sky
coverage is much larger than the large-scale structure  in question.
In addition, it requires that the  variation in survey depth  be
small across  the sky coverage (Li et al.   2006).

Although Li  et al. (2006) found that  two-point correlation functions
based on the random samples  described above match well those obtained
using random samples based on the luminosity function, we need to test
whether  or  not  the  use   of  different  random  samples  can  have
significant  impact on  our results  for the  smaller  subsamples used
here.  To do  this, we generate another set of  random samples for all
galaxies according  to the luminosity function obtained  by Blanton et
al.  (2003b).  Here  a large  set of  points are  randomly distributed
within the survey  sky coverage and each point  is assigned a redshift
and  a magnitude  according to  the luminosity  function. This  set of
random samples will be referred to as type R2.

For both sets of random samples, the {\it mangle} software provided by
A.  Hamilton (Hamilton \& Tegmark  2004) is used to locate the polygon
ID  for each  random  galaxy  and to  determine  the completeness  and
magnitude limit  (which changes slightly across the  sky coverage) for
this polygon  using the mask provided  by Blanton et  al. (2005).  The
completeness and magnitude limit so  obtained are then included in the
random samples. In calculating the two-point correlation functions for
all galaxies  we use random samples  that are 10 time  bigger than the
observational  data. While  for red  or  blue galaxies  we use  random
samples that are 15 times bigger than the observational data.

\section{Method to Estimate the Two-Point Correlation Function}
\label{sec_method}

\begin{figure}
  \plotone{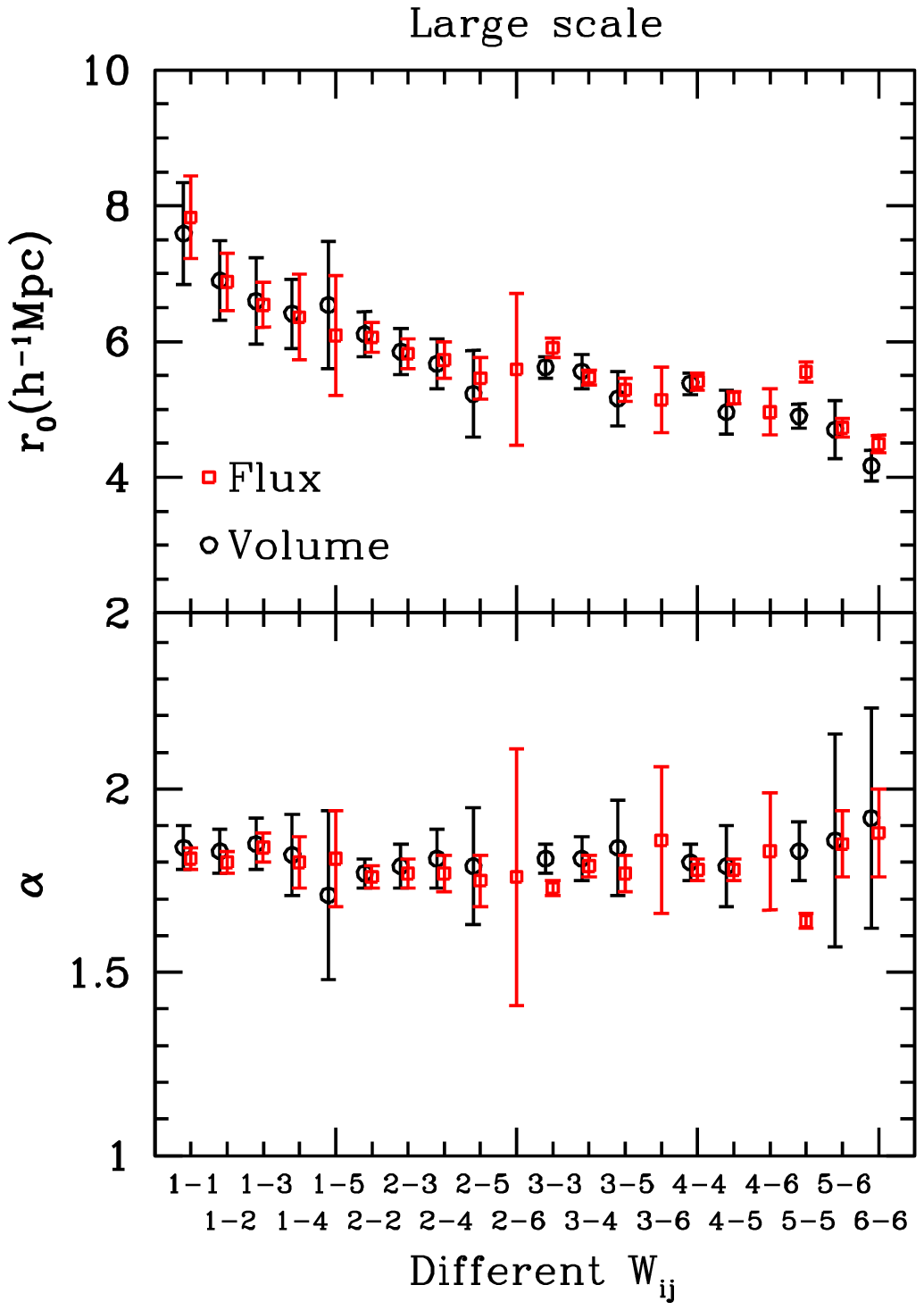}   \caption{The  correlation   length,   $r_0$,  and
    logarithmic slope, $\alpha$, obtained from the projected auto- and
    cross-correlation  functions on  scales $0.98\mpch  \leq  r_p \leq
    9.6\mpch$ for various volume-limited and flux-limited samples. The
    error-bars are  1-$\sigma$ scatter among  the three SDSS  regions.
    In the horizontal axis we  indicate the IDs ($i-j$) of the samples
    used in  the estimate of the correlation  function, $W_{i,j}$. The
    circles  with  error-bars   are  the  results  for  volume-limited
    samples.  The   squares  with  error-bars  are   the  results  for
    flux-limited samples.  Note that since  the overlap region  is too
    small, we  do not have measurement  of 1-6, 2-6, 3-6  and 4-6 data
    points  for  volume-limited   samples  and  1-6  for  flux-limited
    samples.  Overall, the results for volume-limited and flux-limited
    samples  agree  well.   Large  discrepancy exists  for  $W_{5,5}$,
    likely due to the existence of the supercluster at redshift $z\sim
    0.08$.}
\label{fig2}
\end{figure}

\begin{figure}
  \plotone{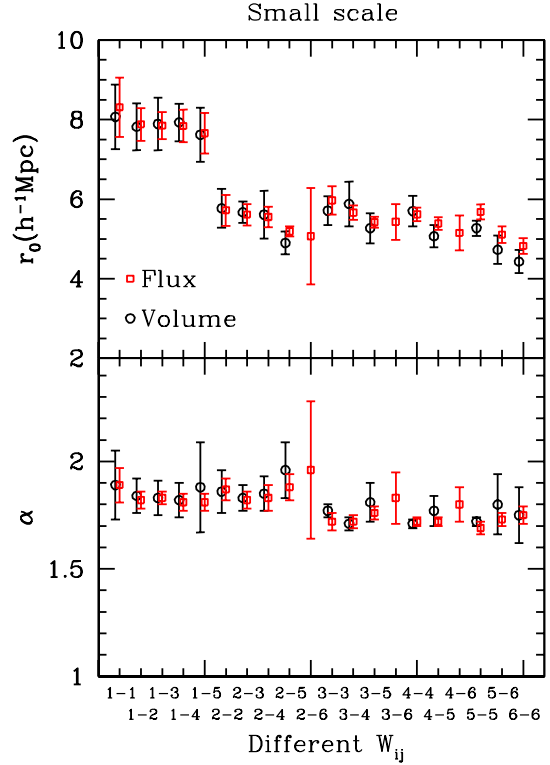}
\caption{The same as Fig.~\ref{fig2} except that the results
  are obtained from the fitting  to the correlation functions on small
  scales ($0.16\mpch \leq r_p  \leq 0.98\mpch$). Note again that since
  the overlap region is too small,  we do not have measurement of 1-6,
  2-6, 3-6 and 4-6 data  points for volume-limited samples and 1-6 for
  flux-limited samples.}
\label{fig3}
\end{figure}

\begin{deluxetable*}{lcccccc}
  \tablecaption{The correlation lengths and slopes for volume-limited
samples on {\bf large} scales ($0.98\mpch \leq r_p \leq 9.6\mpch$)
      \label{tab3}. }
\tablewidth{0pt} \tablehead{ Sample & Va & Vb
    & Vc & Vd & Ve & Vf }

\startdata
& $\alpha$ & $\alpha$  & $\alpha$  & $\alpha$  & $\alpha$ & $\alpha$   \\
& $r_0$    & $r_0$     & $r_0$     & $r_0$     & $r_0$    & $r_0$   \\
\tableline
V1 & $1.84 \pm 0.06$ & & & & & \\
   & $7.59 \pm 0.75$ & & & & & \\

V2 & $1.83 \pm 0.06$ &$1.77 \pm 0.04$ & & & & \\
   & $6.90 \pm 0.59$ &$6.11 \pm 0.33$ & & & & \\

V3 & $1.85 \pm 0.07$ &$1.79 \pm 0.06$ &$1.81 \pm 0.04$ & & & \\
   & $6.60 \pm 0.64$ &$5.85 \pm 0.34$ &$5.62 \pm 0.16$ & & & \\

V4 & $1.82 \pm 0.11$ &$1.81 \pm 0.08$ &$1.81 \pm 0.06$ &$1.80 \pm 0.05$ & & \\
   & $6.41 \pm 0.51$ &$5.67 \pm 0.37$ &$5.56 \pm 0.25$ &$5.38 \pm 0.16$ & & \\

V5 & $1.71 \pm 0.23$ &$1.79 \pm 0.16$ &$1.84 \pm 0.13$ &$1.79 \pm 0.11$ &$1.83 \pm 0.08$ & \\
   & $6.54 \pm 0.94$ &$5.23 \pm 0.64$ &$5.16 \pm 0.40$ &$4.96 \pm 0.32$ &$4.90 \pm 0.18$ & \\

V6 & & & & &$1.86 \pm 0.29$ &$1.92 \pm 0.30$\\
   & & & & &$4.70 \pm 0.43$ &$4.17 \pm 0.23$\\

\tableline \tableline

$\alpha=\overline{\alpha}=1.82$ & $r_0$ & $r_0$ & $r_0$ & $r_0$ & $r_0$ & $r_0$  \\

\tableline

V1& $7.57 \pm 0.45$ & & & & & \\

V2& $6.89 \pm 0.34$ &$6.13 \pm 0.18$ & & & & \\

V3& $6.56 \pm 0.38$ &$5.87 \pm 0.20$ &$5.62 \pm 0.11$ & & & \\

V4& $6.40 \pm 0.39$ &$5.67 \pm 0.23$ &$5.56 \pm 0.14$ &$5.39 \pm 0.11$ & & \\

V5& $6.56 \pm 0.58$ &$5.26 \pm 0.37$ &$5.14 \pm 0.29$ &$4.97 \pm 0.21$ &$4.89 \pm 0.09$ & \\

V6& & & & &$4.65 \pm 0.38$ &$4.08 \pm 0.20$

\enddata \tablecomments{The  upper part lists  the best fit  values of
  $r_0$ and $\alpha$ for  different volume-limited samples, using data
  points  on large scales,  $0.98\mpch \leq  r_p \leq  9.6\mpch$.  The
  table is  not filled up because  the overlaps in  volume between the
  faintest  and  bright  samples  are  too  small  to  allow  reliable
  estimates of the cross correlation functions.  The lower part of the
  table shows the correlation length  obtained by setting the slope of
  the correlation functions, $\alpha$, to be $\overline{\alpha}=1.82$,
  the average of the values listed in the upper part.}
\end{deluxetable*}

\begin{deluxetable*}{lcccccc}
  \tablecaption{The correlation lengths and slopes for volume-limited
  samples on {\bf small} scales ($0.16\mpch \leq r_p \leq 0.98\mpch$)
  \label{tab4}}

\tablewidth{0pt} \tablehead{ Sample & Va & Vb
    & Vc & Vd & Ve & Vf }

\startdata
& $\alpha$ & $\alpha$  & $\alpha$  & $\alpha$  & $\alpha$ & $\alpha$   \\
& $r_0$    & $r_0$     & $r_0$     & $r_0$     & $r_0$    & $r_0$   \\
\tableline

V1 & $1.89 \pm 0.16$ & & & & & \\
   & $8.07 \pm 0.81$ & & & & & \\

V2 & $1.84 \pm 0.08$ &$1.86 \pm 0.10$ & & & & \\
   & $7.82 \pm 0.59$ &$5.77 \pm 0.49$ & & & & \\

V3 & $1.83 \pm 0.08$ &$1.83 \pm 0.06$ &$1.77 \pm 0.03$ & & & \\
   & $7.89 \pm 0.66$ &$5.67 \pm 0.27$ &$5.71 \pm 0.36$ & & & \\

V4 & $1.82 \pm 0.08$ &$1.85 \pm 0.08$ &$1.71 \pm 0.03$ &$1.71 \pm 0.02$ & & \\
   & $7.93 \pm 0.47$ &$5.61 \pm 0.60$ &$5.88 \pm 0.56$ &$5.70 \pm 0.38$ & & \\

V5 & $1.88 \pm 0.21$ &$1.96 \pm 0.13$ &$1.81 \pm 0.09$ &$1.77 \pm 0.07$ &$1.72 \pm 0.02$ & \\
   & $7.62 \pm 0.68$ &$4.90 \pm 0.29$ &$5.27 \pm 0.38$ &$5.07 \pm 0.28$ &$5.27 \pm 0.19$ & \\

V6 & & & & &$1.80 \pm 0.14$ &$1.75 \pm 0.13$\\
   & & & & &$4.73 \pm 0.36$ &$4.43 \pm 0.29$\\

\tableline \tableline

$\alpha=\overline{\alpha}=1.81$ & $r_0$ & $r_0$ & $r_0$ & $r_0$ & $r_0$ & $r_0$  \\

\tableline
V1& $8.16 \pm 0.54$ & & & & & \\

V2& $7.21 \pm 0.40$ &$6.00 \pm 0.13$ & & & & \\

V3& $7.07 \pm 0.36$ &$5.74 \pm 0.12$ &$5.53 \pm 0.09$ & & & \\

V4& $6.99 \pm 0.46$ &$5.76 \pm 0.13$ &$5.44 \pm 0.08$ &$5.27 \pm 0.06$ & & \\

V5& $6.70 \pm 0.26$ &$5.45 \pm 0.23$ &$5.25 \pm 0.12$ &$4.93 \pm 0.09$ &$4.90 \pm 0.11$ & \\

V6& & & & &$4.69 \pm 0.14$ &$4.23 \pm 0.09$

\enddata \tablecomments{The same  as Table~\ref{tab3}, except that the
  results are  obtained by fitting the correlation  functions on small
  scales, $0.16\mpch \leq r_p \leq 0.98\mpch$.}
\end{deluxetable*}

We  estimate both  the 2-point  auto- and  cross-correlation functions
(hereafter 2PCFs)  using the definition  proposed in Davis  \& Peebles
(1983),
\begin{equation}\label{eq:xi}
\xi_{i,j}(r_p,r_\pi)={D_iD_j (r_p, r_\pi)
\over D_iR_j (r_p, r_\pi)}-1 \,,
\end{equation}
where $i,j$ denote the pair of samples for which the cross correlation
function is estimated; $i=j$ corresponds to auto-correlation function. As a
convention, sample $i$ always represents the brighter sample if $i\ne j$.  In
the above definition, $D_iD_j$ is the count of galaxy pairs between samples
$i$ and $j$, $D_iR_j$ is the count of galaxy-random pairs between galaxy
sample $i$ and random sample $j$, and $r_p$ and $r_\pi$ are the separations of
galaxy pairs perpendicular and parallel to the line of sight, respectively.
Galaxy-galaxy and galaxy-random pairs are counted in logarithmic bins in
$r_p$, with bin width $\Delta\log_{10}(r_p)=0.2$, and in linear bins of
$r_\pi$, with bin width $\Delta r_\pi=1\mpch$.  Since we have galaxy samples
in three different regions (NGCE, NGCO and SGC), we combine the number counts
of all regions when estimating the correlation function. The scatter among the
three regions are used to indicate the uncertainties (error-bars) in the
measurements. Note that the errors of the 2PCFs for the faintest sample may be
somewhat underestimated compared to the true cosmic variance because the
volumes and separations of these three regions are too small to represent the
true cosmic variance.

As  mentioned  in  Section~\ref{sec:data_sdss}, the  redshift
overlap between some  of the volume-limited samples is  very small,
especially between the bright  and faint samples.  In fact, samples
V1 and V5 as well   as  V1   and   V6  have   no   overlap-volume at
all  (cf. Fig.\ref{fig:mgr}).  In  those cases we have  to use
samples  Va -- Vf (which  may be  incomplete) to  get better
statistics.   According to Eq.~(\ref{eq:xi}),  one may  circumvent
the problems  to some  extent using combinations  of complete and
incomplete  samples.  For example, one can choose the $i$ sample
from Va - Vf, and the $j$ sample and the corresponding random sample
from V1 -  V6. Since V1 - V6 are complete, this reduces the effects
due to the  incompleteness in samples Va -- Vf.

In  our following  discussion, we  will focus  on the  {\it projected}
2PCF, which is defined as
\begin{equation}\label{eq:W_ij}
W_{i,j}(r_p)= 2\int_0^{\infty} \xi_{i,j} (r_p,r_\pi)d r_\pi = 2 \sum_{k}
\xi_{i,j} (r_p, r_{\pi,k}) \Delta r_{\pi, k} \,.
\end{equation}
Here the summation  is made over $k=1$ to  40 corresponding to $r_\pi=
0.5 \mpch$  to $r_\pi = 39.5\mpch$.  If we assume  that the real-space
2PCF $\xi(r)$ can be fitted by a power law,
\begin{equation}
\xi(r)=(r_0/r)^{\alpha}\,,
\end{equation}
where $r_0$ and $\alpha$ are, respectively, the correlation length and
slope of the correlation function,  then the projected 2PCF is related
to the real-space 2PCF as
\begin{equation}\label{eq:W_r_p}
W(r_p) = r_0^\alpha r_p^{1-\alpha}\frac{\Gamma (1/2) \Gamma [(\alpha
-1)/2]}{\Gamma (\alpha/2)}\,,
\end{equation}
where $\Gamma (x)$  is the Gamma function.  As we  will see below, the
observed 2PCFs  can all be fitted  reasonably well by  power laws over
limited  ranges of  $r_p$.  We can  then  use $r_0$  to represent  the
correlation  amplitude,  and $\alpha$  the  shape  of the  correlation
function.

\section{Dependence on galaxy luminosity}
\label{sec:result_L}

\begin{figure*}
\plotone{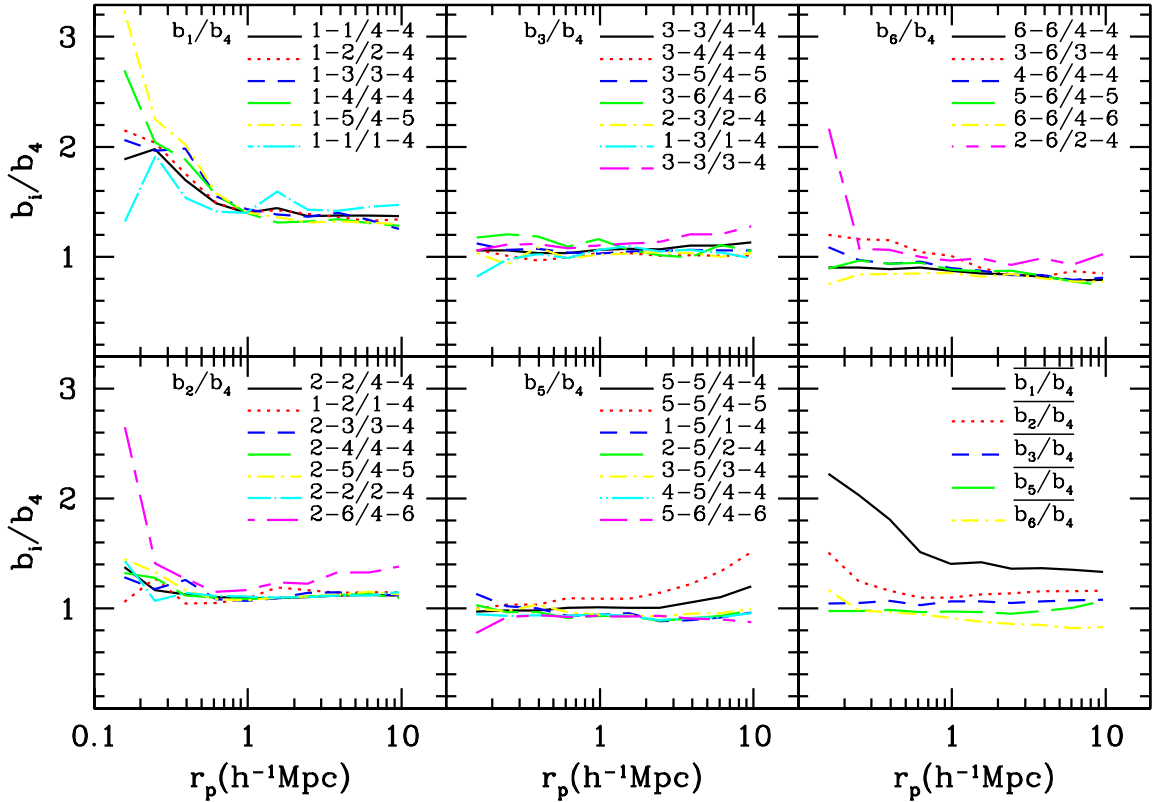}
  \caption{The relative bias as a function of $r_p$ obtained from
    the  ratios   of  the  projected  correlation   functions  of  the
    flux-limited samples.  In the  lower-right panel, we show the mean
    relative bias  parameters obtained by averaging  the results shown
    in each  of the five panels.  In each panel, $i-j$  means that the
    projected  two-point correlation  function is  for samples  Fi and
    Fj.}
\label{fig4}
\end{figure*}

\begin{figure}
  \plotone{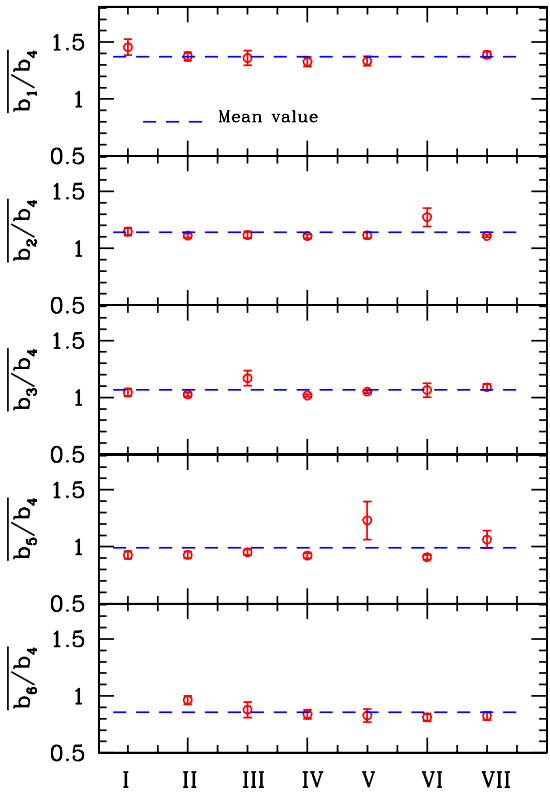} \caption{The relative bias parameters obtained from
    the  correlation functions  over $0.98\mpch$  to  $9.8\mpch$ using
    seven different  approaches (see the main text  for details).  The
    error-bars  are based on  the scatter  of the  data points  in the
    range  $0.98\mpch< r_p\le  9.8\mpch$.  Panels  from top  to bottom
    correspond  to   the  bias  parameters   for  Fi  ($i=1,2,3,5,6$),
    normalized by that of F4. The horizontal axis labels the different
    approaches that are used to  determine the bias parameter for each
    subsample.   One data  point is  missing  in the  first and  fifth
    panels,  because the  redshift overlap  between F1  and F6  is too
    small to measure the cross-correlation function reliably.}
\label{fig5}
\end{figure}

\begin{figure*}
  \plotone{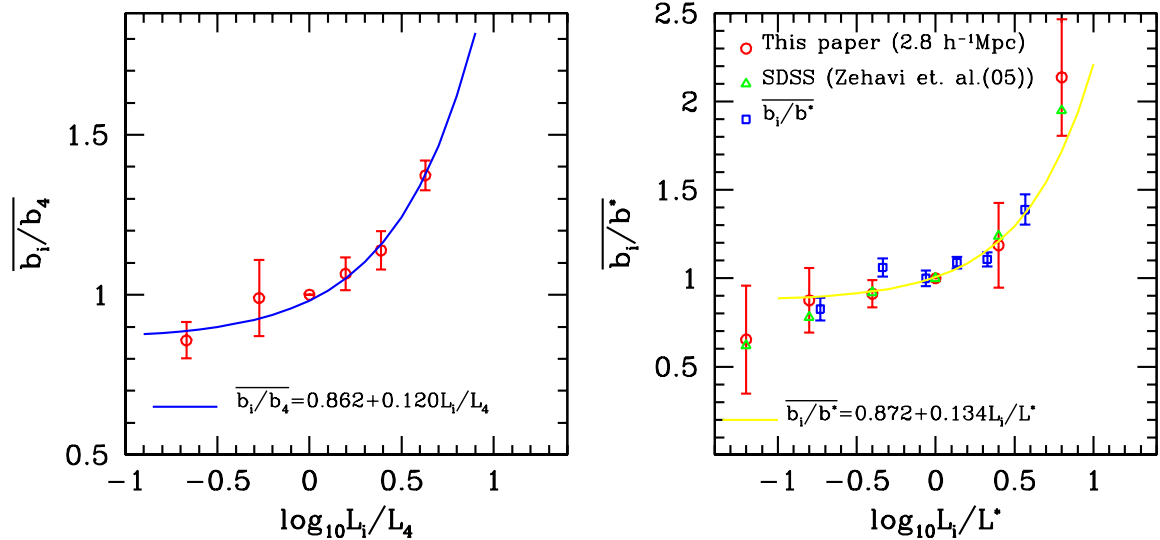}  \caption{Left  panel:  The average  relative  bias
    parameter,    obtained    using    the   correlation    data    in
    $0.98\mpch<r_p\le 9.8\mpch$  as a  function of galaxy  luminosity.
    The results are obtained  using the flux-limited samples, and both
    the bias parameter and the luminosity are normalized by the values
    for F4. For  a given luminosity, the mean  and error-bar are based
    on the different measurements  shown in Fig~\ref{fig5}.  The solid
    curve  shows the relation,  $\overline{ b_i/b_4  }= 0.862  + 0.120
    L_i/L_4$,  obtained by  fitting the  data.  Right  panel:  In this
    panel, we compare our results with those obtained by Zehavi et al.
    (2005).  Here  the  bias   parameters  and  the  luminosities  are
    normalized by  the values for  $L^*$ galaxies.  In Zehavi  et al.,
    the  relative   bias  parameters  are  measured   using  the  {\it
      auto}-correlation   functions    at   $r_p=2.7\mpch$   for   six
    volume-limited  samples   with  luminosities  in  $(-23.0,-22.0)$,
    $(-22.0,-21.0)$,         $(-21.0,-20.0)$,         $(-20.0,-19.0)$,
    $(-19.0,-18.0)$, $(-18.0,-17.0)$,  and their results  are shown by
    triangles.  Using the same  method and  same samples,  our results
    (however at  $r_p=2.8\mpch$) are shown  as open circles  and match
    very well their  results.  The open squares show  the same results
    as shown  in the  left panel, except  that the  normalizations are
    different.  The curve  is a fit to the  results represented by the
    open squares.}
\label{fig6}
\end{figure*}

\begin{figure*}
  \plotone{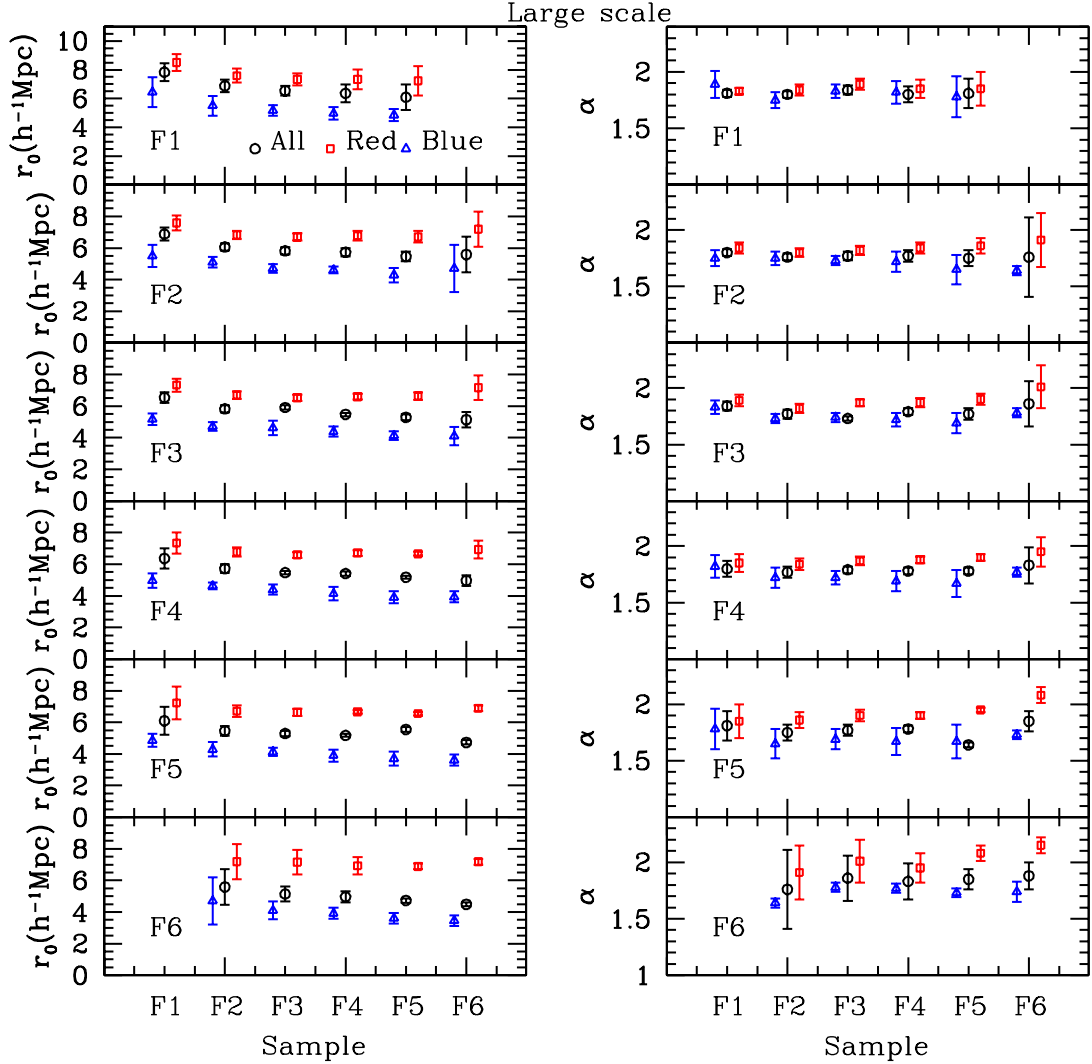}  \caption{Here we  compare the  correlation lengths
    ($r_0$) and slopes ($\alpha$)  of all (circles), red (squares) and
    blue (triangles) galaxies using flux-limited samples (F1 -- F6 for
    all  galaxies;  $\FIred$  --   $\FVIred$  for  red  galaxies,  and
    $\FIblue$ -- $\FVIblue$ for  blue galaxies).  Here the results are
    based on  correlation functions  on large scales,  $0.98\mpch \leq
    r_p \leq  9.6\mpch$. To avoid  confusion, data points for  rad and
    blue galaxies are shifted slightly horizontally.} \label{fig7}
\end{figure*}
\begin{figure*}
  \plotone{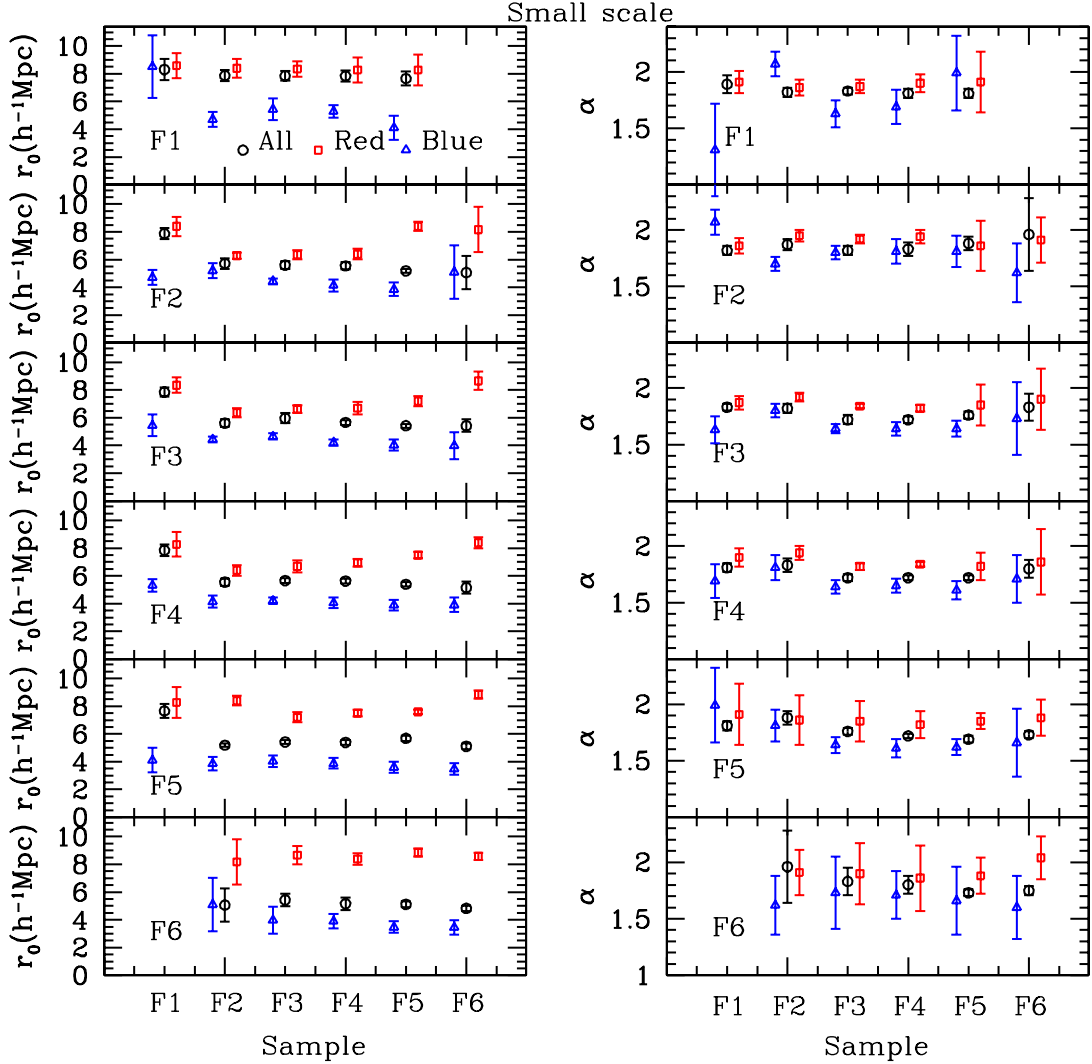} \caption{The  same as Fig.~\ref{fig7},  except that
    the  results  are based  on  the  correlation  functions on  small
    scales, $0.16\mpch \leq r_p \leq 0.98\mpch$.}
\label{fig8}
\end{figure*}
\begin{figure*}
  \plotone{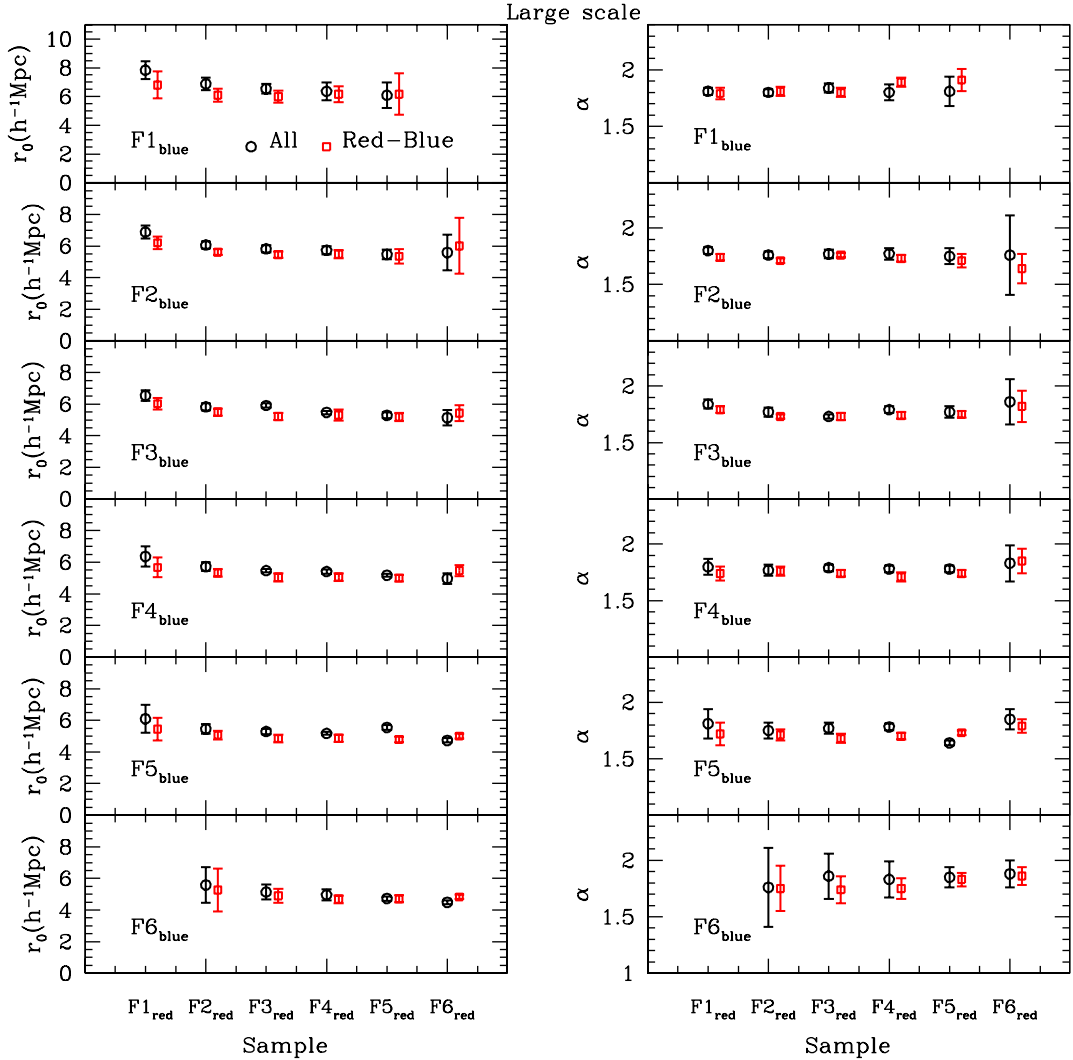}
  \caption{The correlation lengths ($r_0$) and slopes
    ($\alpha$) of the cross-correlation functions between red and blue
    galaxies  (squares)  on large  scales,  $0.98\mpch  \leq r_p  \leq
    9.6\mpch$.  Panels  from top to bottom correspond  to flux limited
    samples  for blue galaxies  ${\rm Fi_{blue}}$  ($i=1,2...,6$). The
    red  samples used  in the  cross-correlations are  labeled  on the
    horizontal axis ${\rm  Fi_{red}}$ ($i=1,2...,6$).  For comparison,
    results for all galaxies in the corresponding flux-limited samples
    are plotted as open circles.}
  \label{fig9}
\end{figure*}

\begin{figure*}
  \plotone{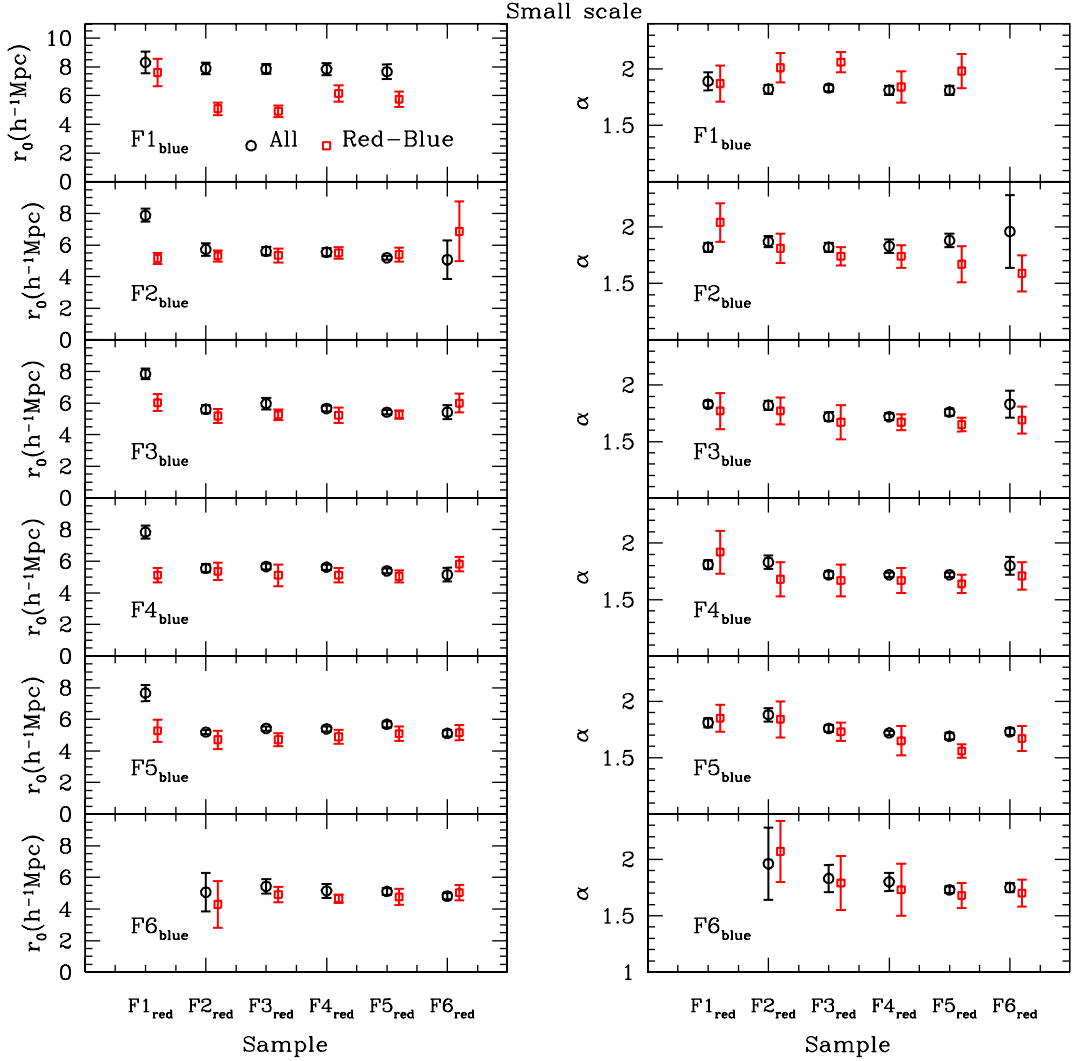}
  \caption{The same as Fig.~\ref{fig9}, but here results
    are  based on  correlation functions  on small  scales, $0.16\mpch
    \leq r_p \leq 0.98\mpch$. }
  \label{fig10}
\end{figure*}

\begin{deluxetable*}{lcccccc}
  \tablecaption{The correlation lengths and slopes for flux-limited
    samples on large scales ($0.98\mpch \leq r_p \leq 9.6\mpch$)
\label{tab5}}
\tablewidth{0pt} \tablehead{ Sample & F1 & F2 & F3 &
    F4 & F5 & F6 }

\startdata
& $\alpha$ & $\alpha$  & $\alpha$  & $\alpha$  & $\alpha$ & $\alpha$   \\
& $r_0$    & $r_0$     & $r_0$     & $r_0$     & $r_0$    & $r_0$   \\

\tableline
F1 & $ 1.81 \pm 0.03$ & & & & & \\
   & $ 7.83 \pm 0.61$ & & & & & \\

F2 & $ 1.80 \pm 0.03$ &$ 1.76 \pm 0.03$ & & & & \\
   & $ 6.88 \pm 0.42$ &$ 6.06 \pm 0.22$ & & & & \\

F3 & $ 1.84 \pm 0.04$ &$ 1.77 \pm 0.04$ &$ 1.73 \pm 0.02$ & & &\\
   & $ 6.54 \pm 0.33$ &$ 5.82 \pm 0.22$ &$ 5.91 \pm 0.14$ & & &\\

F4 & $ 1.80 \pm 0.07$ &$ 1.77 \pm 0.05$ &$ 1.79 \pm 0.03$ &$ 1.78 \pm 0.03$ & &\\
   & $ 6.36 \pm 0.63$ &$ 5.73 \pm 0.27$ &$ 5.47 \pm 0.11$ &$ 5.41 \pm 0.13$ & &\\

F5 & $ 1.81 \pm 0.13$ &$ 1.75 \pm 0.07$ &$ 1.77 \pm 0.05$ &$ 1.78 \pm 0.03$ &$ 1.64 \pm 0.02$ &\\
   & $ 6.09 \pm 0.88$ &$ 5.46 \pm 0.31$ &$ 5.29 \pm 0.17$ &$ 5.17 \pm 0.09$ &$ 5.55 \pm 0.15$ &\\

F6 & &$ 1.76 \pm 0.35$ &$ 1.86 \pm 0.20$ &$ 1.83 \pm 0.16$ &$ 1.85 \pm 0.09$ &$ 1.88 \pm 0.12$\\
   & &$ 5.59 \pm 1.12$ &$ 5.14 \pm 0.48$ &$ 4.96 \pm 0.34$ &$ 4.73 \pm 0.14$ &$ 4.49 \pm 0.13$\\

\tableline \tableline

$\alpha=\overline{\alpha}=1.79$ & $r_0$ & $r_0$ & $r_0$ & $r_0$ & $r_0$ & $r_0$  \\

 \tableline
F1 & $ 7.80 \pm 0.31$ & & & & & \\

F2 & $ 6.86 \pm 0.21$ &$ 6.09 \pm 0.13$ & & & & \\

F3 & $ 6.49 \pm 0.19$ &$ 5.83 \pm 0.12$ &$ 5.94 \pm 0.14$ & & & \\

F4 & $ 6.34 \pm 0.35$ &$ 5.75 \pm 0.18$ &$ 5.48 \pm 0.15$ &$ 5.42 \pm 0.14$ & & \\

F5 & $ 6.07 \pm 0.56$ &$ 5.48 \pm 0.29$ &$ 5.31 \pm 0.10$ &$ 5.17 \pm 0.09$ &$ 5.65 \pm 0.11$ &\\

F6 & &$5.62 \pm 0.46$ &$5.08 \pm 0.37$ &$ 4.93 \pm 0.21$ &$4.68 \pm 0.11$ &$4.38 \pm 0.12$

\enddata
\end{deluxetable*}

\begin{deluxetable*}{lcccccc}
  \tablecaption{The correlation lengths and slopes for flux-limited
    samples on small scales ($0.16\mpch \leq r_p \leq 0.98 \mpch$)
\label{tab6}}
\tablewidth{0pt} \tablehead{ Sample & F1 & F2 & F3 &
    F4 & F5 & F6 }

\startdata
& $\alpha$ & $\alpha$  & $\alpha$  & $\alpha$  & $\alpha$ & $\alpha$   \\
& $r_0$    & $r_0$     & $r_0$     & $r_0$     & $r_0$    & $r_0$   \\
\tableline

F1 & $ 1.89 \pm 0.08$ & & & & &\\
     & $ 8.31 \pm 0.75$ & & & & &\\

F2 & $ 1.82 \pm 0.04$ &$ 1.87 \pm 0.05$ & & & &\\
     & $ 7.88 \pm 0.41$ &$ 5.72 \pm 0.39$ & & & &\\

F3 & $ 1.83 \pm 0.03$ &$ 1.82 \pm 0.04$ &$ 1.72 \pm 0.04$ & & &\\
     & $ 7.85 \pm 0.34$ &$ 5.61 \pm 0.27$ &$ 5.97 \pm 0.36$ & & &\\

F4 & $ 1.81 \pm 0.04$ &$ 1.83 \pm 0.06$ &$ 1.72 \pm 0.03$ &$ 1.72 \pm 0.02$ & &\\
     & $ 7.84 \pm 0.41$ &$ 5.55 \pm 0.26$ &$ 5.66 \pm 0.18$ &$ 5.62 \pm 0.17$ & &\\

F5 & $ 1.81 \pm 0.04$ &$ 1.88 \pm 0.06$ &$ 1.76 \pm 0.03$ &$ 1.72 \pm 0.02$ &$ 1.69 \pm 0.03$ &\\
     & $ 7.66 \pm 0.51$ &$ 5.19 \pm 0.13$ &$ 5.42 \pm 0.14$ &$ 5.39 \pm 0.16$ &$ 5.68 \pm 0.19$ &\\

F6 & &$ 1.96 \pm 0.32$ &$ 1.83 \pm 0.12$ &$ 1.80 \pm 0.08$ &$ 1.73 \pm 0.03$ &$ 1.75 \pm 0.04$\\
     & &$ 5.07 \pm 1.21$ &$ 5.43 \pm 0.45$ &$ 5.15 \pm 0.44$ &$ 5.11 \pm 0.21$ &$4.82 \pm 0.20$\\

\tableline \tableline

$\alpha=\overline{\alpha}=1.80$ & $r_0$ & $r_0$ & $r_0$ & $r_0$ & $r_0$ & $r_0$  \\

 \tableline

F1 & $8.21 \pm 0.42$ & & & & & \\

F2 & $7.17 \pm 0.23$ &$6.01 \pm 0.19$ & & & & \\

F3 & $6.93 \pm 0.11$ &$5.71 \pm 0.19$ &$5.61 \pm 0.15$ & & & \\

F4 & $6.96 \pm 0.13$ &$5.69 \pm 0.18$ &$5.33 \pm 0.06$ &$5.29 \pm 0.15$ & & \\

F5 & $6.77 \pm 0.16$ &$5.50 \pm 0.15$ &$5.25 \pm 0.17$ &$5.10 \pm 0.18$ &$5.24 \pm 0.14$ & \\

F6 & &$5.69 \pm 0.59$ &$5.54 \pm 0.09$ &$5.13 \pm 0.13$ &$4.85 \pm 0.17$ &$4.64 \pm 0.15$

\enddata
\end{deluxetable*}

\begin{deluxetable*}{lcccccc}
  \tablecaption{The correlation lengths and slopes for flux-limited samples of
    red galaxies on large scales ($0.98\mpch \leq r_p \leq 9.6\mpch$)
\label{tab7}}
\tablewidth{0pt}
\tablehead{ Sample
& ${\rm F1_{red}}$ & ${\rm F2_{red}}$ & ${\rm F3_{red}}$
& ${\rm F4_{red}}$ & ${\rm F5_{red}}$ & ${\rm F6_{red}}$ }

\startdata
& $\alpha$ & $\alpha$  & $\alpha$  & $\alpha$  & $\alpha$ & $\alpha$   \\
& $r_0$    & $r_0$     & $r_0$     & $r_0$     & $r_0$    & $r_0$   \\
\tableline

${\rm F1_{red}}$ & $ 1.83 \pm 0.03$ & & & & &\\
    & $ 8.50 \pm 0.59$ & & & & &\\

${\rm F2_{red}}$ & $ 1.84 \pm 0.05$ &$ 1.80 \pm 0.04$ & & & &\\
    & $ 7.59 \pm 0.48$ &$ 6.82 \pm 0.26$ & & & &\\

${\rm F3_{red}}$ & $ 1.89 \pm 0.05$ &$ 1.82 \pm 0.04$ &$ 1.87 \pm 0.03$ & & &\\
    & $ 7.33 \pm 0.41$ &$ 6.70 \pm 0.24$ &$ 6.53 \pm 0.20$ & & &\\

${\rm F4_{red}}$ & $ 1.85 \pm 0.08$ &$ 1.84 \pm 0.05$ &$ 1.87 \pm 0.04$ &$ 1.88 \pm 0.03$ & &\\
    & $ 7.34 \pm 0.68$ &$ 6.78 \pm 0.30$ &$ 6.59 \pm 0.18$ &$ 6.71 \pm 0.18$ & &\\

${\rm F5_{red}}$ & $ 1.85 \pm 0.15$ &$ 1.86 \pm 0.07$ &$ 1.90 \pm 0.05$ &$ 1.90 \pm 0.03$ &$ 1.95 \pm 0.02$ &\\
    & $ 7.23 \pm 1.03$ &$ 6.71 \pm 0.36$ &$ 6.63 \pm 0.24$ &$ 6.66 \pm 0.12$ &$ 6.55 \pm 0.11$ &\\

${\rm F6_{red}}$ & &$ 1.91 \pm 0.24$ &$ 2.01 \pm 0.19$ &$ 1.95 \pm 0.13$ &$ 2.08 \pm 0.07$ &$ 2.15 \pm 0.07$\\
    & &$ 7.19 \pm 1.11$ &$ 7.16 \pm 0.78$ &$ 6.93 \pm 0.56$ &$ 6.89 \pm 0.18$ &$ 7.18 \pm 0.18$\\

\tableline \tableline

$\alpha=\overline{\alpha}=1.90$ & $r_0$ & $r_0$ & $r_0$ & $r_0$ & $r_0$ & $r_0$  \\

 \tableline

${\rm F1_{red}}$ & $8.50 \pm 0.44$ & & & & & \\

${\rm F2_{red}}$ & $7.60 \pm 0.36$ &$6.82 \pm 0.25$ & & & & \\

${\rm F3_{red}}$ & $7.33 \pm 0.35$ &$6.71 \pm 0.25$ &$6.52 \pm 0.13$ & & & \\

${\rm F4_{red}}$ & $7.35 \pm 0.37$ &$6.77 \pm 0.34$ &$6.59 \pm 0.21$ &$6.70 \pm 0.14$ & & \\

${\rm F5_{red}}$ & $7.24 \pm 0.32$ &$6.70 \pm 0.33$ &$6.62 \pm 0.23$ &$6.66 \pm 0.12$ &$6.56 \pm 0.09$ & \\

${\rm F6_{red}}$ & &$7.19 \pm 0.78$ &$7.17 \pm 0.19$ &$6.92 \pm
0.22$ &$6.97 \pm 0.29$ &$7.25 \pm 0.31$

\enddata

\end{deluxetable*}

\begin{deluxetable*}{lcccccc}
  \tablecaption{The correlation lengths and slopes for flux-limited samples of
    red galaxies on small scales ($0.16\mpch \leq r_p \leq 0.98\mpch$)
\label{tab8}}
\tablewidth{0pt} \tablehead{Sample &
   ${\rm F1_{red}}$ & ${\rm F2_{red}}$ & ${\rm F3_{red}}$ & ${\rm F4_{red}}$ &
   ${\rm F5_{red}}$ & ${\rm F6_{red}}$ }

\startdata
& $\alpha$ & $\alpha$  & $\alpha$  & $\alpha$  & $\alpha$ & $\alpha$   \\
& $r_0$    & $r_0$     & $r_0$     & $r_0$     & $r_0$    & $r_0$   \\
\tableline

${\rm F1_{red}}$ & $ 1.91 \pm 0.10$ & & & & &\\
    & $ 8.59 \pm 0.90$ & & & & &\\

${\rm F2_{red}}$ & $ 1.86 \pm 0.07$ &$ 1.95 \pm 0.05$ & & & &\\
    & $ 8.40 \pm 0.69$ &$ 6.28 \pm 0.23$ & & & &\\

${\rm F3_{red}}$ & $ 1.87 \pm 0.06$ &$ 1.92 \pm 0.04$ &$ 1.84 \pm 0.02$ & & &\\
    & $ 8.35 \pm 0.55$ &$ 6.36 \pm 0.34$ &$ 6.62 \pm 0.28$ & & &\\

${\rm F4_{red}}$ & $ 1.90 \pm 0.08$ &$ 1.94 \pm 0.06$ &$ 1.82 \pm 0.03$ &$ 1.84 \pm 0.02$ & &\\
    & $ 8.27 \pm 0.89$ &$ 6.40 \pm 0.38$ &$ 6.69 \pm 0.44$ &$ 6.94 \pm 0.27$ & &\\

${\rm F5_{red}}$ & $ 1.91 \pm 0.27$ &$ 1.86 \pm 0.22$ &$ 1.85 \pm 0.18$ &$ 1.82 \pm 0.12$ &$ 1.85 \pm 0.07$ &\\
    & $ 8.28 \pm 1.12$ &$ 8.40 \pm 0.34$ &$ 7.20 \pm 0.36$ &$ 7.50 \pm 0.24$ &$ 7.59 \pm 0.16$ &\\

${\rm F6_{red}}$ & &$ 1.91 \pm 0.20$ &$ 1.90 \pm 0.27$ &$ 1.86 \pm 0.29$ &$ 1.88 \pm 0.16$ &$ 2.04 \pm 0.19$\\
    & &$ 8.17 \pm 1.64$ &$ 8.66 \pm 0.66$ &$ 8.39 \pm 0.41$ &$ 8.85 \pm 0.29$ &$ 8.58 \pm 0.26$\\

\tableline \tableline

$\alpha=\overline{\alpha}=1.89$ & $r_0$ & $r_0$ & $r_0$ & $r_0$ & $r_0$ & $r_0$  \\

 \tableline

${\rm F1_{red}}$ & $8.52 \pm 0.49$ & & & & & \\

${\rm F2_{red}}$ & $7.62 \pm 0.43$ &$6.61 \pm 0.18$ & & & & \\

${\rm F3_{red}}$ & $7.75 \pm 0.33$ &$6.54 \pm 0.14$ &$6.42 \pm 0.31$ & & & \\

${\rm F4_{red}}$ & $8.04 \pm 0.23$ &$6.72 \pm 0.16$ &$6.59 \pm 0.33$ &$6.76 \pm 0.26$ & & \\

${\rm F5_{red}}$ & $8.13 \pm 0.49$ &$7.03 \pm 0.18$ &$6.97 \pm 0.16$ &$7.00 \pm 0.24$ &$7.33 \pm 0.15$ & \\

${\rm F6_{red}}$ &  &$7.63 \pm 0.83$ &$8.78 \pm 0.36$ &$8.04 \pm 0.34$
&$8.64 \pm 0.22$ &$10.12 \pm 0.25$

\enddata
\end{deluxetable*}

We start  our investigation by comparing the  projected 2PCFs obtained
using the  volume-limited and flux-limited  samples.  When calculating
the  projected   2PCFs  with   the  volume-limited  samples,   we  use
combinations of  the set  V1--V6 with the  set Va--Vf for  the reasons
given  in  Section  ~\ref{sec_method}.   Since the  results  are  very
similar for  the volume-limited and the flux-limited  samples, we show
only  the results  for  the flux-limited  samples in  Fig.~\ref{auto},
where we compare  the projected 2PCFs based on  random samples of type
R1 with those based  on random samples of type R2.  As  found by Li et
al.  (2006),  the two  types of random  samples give  almost identical
results in  most cases.  The largest  discrepancy is seen  in the auto
2PCF  of sample  F5,  This  sample is  significantly  affected by  the
supercluter at redshift  $z\sim 0.08$, and the points  in sample R1 is
not completely random due to the way it is constructed.  Since overall
the  random  samples  of type  R1  work  quite  well and  since  their
construction does not require the luminosity function of galaxies, the
results presented below will be based  on this type of random samples.
A comparison  of the various projected 2PCFs  in Fig.~\ref{auto} shows
that  brighter galaxies  are on  average more  strongly  clustered, an
effect that we will describe in more detail in the following.

To  quantify  the  observed  2PCFs,  we fit  to  the  projected  2PCFs
according to  Eq.~(\ref{eq:W_r_p}) with the two  free parameters $r_0$
and $\alpha$.  A  larger value of $r_0$ implies  a stronger clustering
strength,  while  a  larger  $\alpha$ implies  a  steeper  correlation
function. In  order to  see how $r_0$  and $\alpha$ might  change with
scales, we fit  the projected 2PCFs separately over  two scale ranges,
$0.98\mpch  \leq  r_p \leq  9.6\mpch$  and  $0.16\mpch  \leq r_p  \leq
0.98\mpch$. The separation  between large and small scales  is made at
$r_p\sim 1\mpch$,  because the correlation function  on smaller scales
is expected to be dominated by the one-halo term, while that on larger
scales by the  two-halo term, as shown in Yang et  al. (2005).  We use
two methods to obtain $r_0$ and  $\alpha$ from the data.  In the first
method,  we obtain $r_0$  and $\alpha$  directly from  the fit  to the
projected 2PCFs. In  the second one, we first  obtain an average value
of $\alpha$,  ${\overline\alpha}$, from all  the correlation functions
and   then   fit   the   projected   2PCF   to   get   $r_0$   keeping
$\alpha={\overline\alpha}$.   The fitting  parameters so  obtained are
listed  in Tables~\ref{tab3},  \ref{tab4}, \ref{tab5}  and \ref{tab6},
separately for volume-limited and  flux-limited samples, and for large
and small scales.

Fig.~\ref{fig2}  compares the values  of $r_0$  and $\alpha$  on large
scales  ($0.98\mpch  \leq  r_p   \leq  9.6\mpch$)  obtained  from  the
volume-limited  samples  with those  obtained  from the  corresponding
flux-limited  samples. Overall,  the results  for these  two  kinds of
samples are similar. The only exception is $W_{5,5}$, where the values
of $r_0$ and $\alpha$  for the volume-limited and flux-limited samples
are  significantly  different, largely  because  the  presence of  the
supercluster  at $z\sim  0.08$ that  makes the  correlation properties
quite sensitive to how the sample  is exactly formed. There is a clear
trend  in the  correlation length  $r_0$ that  brighter  galaxies have
larger $r_0$. The cross-correlations between bright and faint galaxies
are also stronger than the auto correlations of faint galaxies. On the
other hand  the slope  of the correlation  function does not  show any
strong trend  with luminosity,  and the mean  value of and  scatter in
$\alpha$  are  $1.82\pm  0.04$  for  the  volume-limited  samples  and
$1.79\pm 0.05$  for the flux-limited samples.  The correlation lengths
obtained by  fixing $\alpha$ at these  mean values are  also listed in
Tables   ~\ref{tab3},  \ref{tab4},   \ref{tab5}  and   \ref{tab6}  for
comparison.

In order to  investigate how the clustering strengths  change on small
scales,  we fit  the  projected 2PCFs  with  power laws  in the  range
$0.16\mpch  \leq r_p  \leq 0.98\mpch$,  and the  results are  shown in
Fig.~\ref{fig3}  and listed  in Tables~\ref{tab4}  and  \ref{tab6}.  A
comparison  to the  results shown  in Fig.~\ref{fig2}  shows  that the
clustering  strengths  for  the  samples that  involve  the  brightest
galaxies  are  enhanced  on  small  scales; in  particular  the  cross
correlation  between  the  brightest  and faintest  galaxies  is  much
stronger  on  small  scales.    For  faint  galaxies,  the  clustering
strengths on small scales are similar to those based on the power laws
obtained on  large scales. This  suggests that there is  a significant
population of faint galaxies that are distributed around the brightest
galaxies  and  the number  drops  rapidly  with  the increase  of  the
distance  to the  brightest galaxies.   As we  will see  below,  it is
likely  that this  population is  dominated by  faint red  galaxies in
clusters  and groups.

Having investigated the clustering strengths for galaxies of different
luminosities,  we now  study the  bias factor  in the  distribution of
galaxies as a function of galaxy luminosity.  Since we do not know the
two-point  correlation function of  the cosmic  density field,  we can
only study the bias factor in  a relative sense. To do this, we choose
F4 as  our fiducial sample,  because it has  a good overlap  in volume
with both the faint and bright samples. Since each sample is used in a
number of projected 2PCFs, we  can estimate the relative bias for each
sample in a number of ways.  With F4 as the reference sample, the bias
factor for the $i$th sample can be estimated from one of the following
definitions:
\begin{eqnarray}\label{eq:bi_b4}
\frac{b_i}{b_4} &=& \frac{W_{i,j}}{W_{j,4}} (j=1, 2, 3, 4, 5, 6) \,,\\
\nonumber
\frac{b_i}{b_4} &=& \sqrt{\frac{W_{i,i}}{W_{4,4}}}\,,
\end{eqnarray}
where $i=1,2,3,5,6$.  Thus, each relative bias can  be estimated using
seven different  approaches. In  what follows, these  seven approaches
will  be   referred  as  approaches  I,   II,  $\cdot\cdot\cdot$,  VI,
corresponding   to  $j=1,2,\cdot\cdot\cdot,   6$,   respectively,  and
approach VII,  corresponding to the  definition in the second  line of
equation  (\ref{eq:bi_b4}).  Note  that in  general the  relative bias
factors,  $b_i/b_4$,  obtained   from  different  approaches  are  not
expected   to  be   the  same.   As   we  will   discuss  in   Section
\ref{sec:discussion}, these  relative bias factors are  expected to be
the  same only  under the  assumption that  the bias  for  galaxies of
different luminosities is linear  and that the stochastic component in
the bias relations between different galaxies is small.

The first 5 panels in Fig.~\ref{fig4} show the relative bias factor as
a function of $r_p$ for  the five samples, Fi ($i=1,2,3,5,6$), and the
7 different  lines within each  panel show the results  obtained using
the 7 different  approaches described above. For most  cases, the bias
function is  almost independent of $r_p$ at  $r_p> 1\mpch$, indicating
that nonlinearity in  the relative bias is small  on these scales. The
bias  factors obtained using  the 7  different approaches  agree quite
well  with each  other. Their  averages are  shown in  the lower-right
panel which clearly shows  that brighter galaxies have higher relative
biases.

To quantify the relative bias as a function of luminosity, we estimate
a  mean relative  bias  factor for  each  case by  averaging the  bias
function  over   $1\mpch$-$10  \mpch$.   The  results  are   shown  in
Fig.~\ref{fig5}   for  different  luminosity   samples  using   the  7
approaches.  Note  that there is one  data point missing  in the first
and fifth panels, because the overlap between samples F1 and F6 is too
small to have a reliable  measurement of $W_{1,6}$. The dashed line in
each panel is the average value of the relative bias obtained from the
seven different approaches. As one  can see, the relative bias factors
obtained with different approaches  are consistent with each other. We
estimate  the mean  luminosity of  galaxies in  each sample  F1--F6 as
$L_i$, and the corresponding average absolute magnitudes thus obtained
are  $-21.87$, $-21.24$, $-20.76$,  $-20.27$, $-19.61$,  and $-18.62$,
respectively.  These  magnitudes are then  converted into luminosities
to  represent  the  characteristic  luminosities  of  the  samples  in
question. The left panel of Fig.~\ref{fig6} shows the average relative
bias  factor  versus  $L_i/L_4$   as  open  circles,  with  error-bars
representing the scatter among  the seven different approaches.  There
is  a  clear  trend  that  the relative  bias  increases  with  galaxy
luminosity.  The   solid  curve   is  a  linear   fit  to   the  data:
$\overline{b_i/b_4} = 0.862 + 0.120 L_i/L_4$.

The  luminosity  dependence  of  galaxy  bias in  the  SDSS  has  been
investigated by Zehavi et al.  (2005) using the amplitudes of the auto
correlation  functions  at  a  fixed projected  radius  $r_p=2.7\mpch$
obtained from  six volume-limited  samples with $M_r$  in $(-23,-22)$,
$(-22,-21)$,  $(-21,-20)$, $(-20,-19)$, $(-19,-18)$,  $(-18,-17)$. The
bias factors were normalized relative  to $L^*$ galaxies defined to be
the  ones with  $M_r$  in  $(-21,-20)$. To  compare  our results  with
theirs, we  show in  the right panel  of Fig.~\ref{fig6}  the relative
bias factors obtained by Zehavi  et al. as triangles together with our
results  (open circles)  that are  obtained from  the auto-correlation
amplitudes  at a  fixed, but  a slightly  different,  projected radius
$r_p=2.8\mpch$. Note that our  results are in excellent agreement with
the results obtained by Zehavi et al.  (2005). For the relative bias -
luminosity relation shown in the left panel of Fig.~\ref{fig6}, we can
convert it into a relation with $L/L^*$. The characteristic luminosity
for sample  F* is $-20.47$, very  close to $L^*$.  Using the projected
2PCF for F*, $W_{\star,\star}$, we can write
\begin{equation}
\overline{b_i/b^*}=\overline{b_i/b_4 \sqrt{\frac{W_{4,4}}{W_{*,*}}}}\,,
\end{equation}
where $W_{4,4}$ and $W_{*,*}$ are the projected auto 2PCFs of samples F4 and
F*, respectively. We first calculate $b_i/b^*$ at different scales and then
average these biases over $1\mpch$-$10 \mpch$ to obtain the average bias
$\overline{b_i/b^*}$. The resulting $\overline{b_i/b^*}$ - $L/L_\star$
relation is shown by the open squares in the right panel of Fig.~\ref{fig6}.
Fitted with a linear model, this relation can be described as
$\overline{b_i/b^*}= 0.872 + 0.134 L_i/L^*$. As one can see, our results based
on the cross-correlation match well those of Zehavi et al.  (2005). The
implications of this agreement will be discussed in Section
\ref{sec:discussion}.

\section{Dependence on galaxy color}
\label{sec:result_c}

\begin{deluxetable*}{lcccccc}
  \tablecaption{The correlation length and slope for flux-limited
    samples of blue galaxies on large scales
    ($0.98\mpch \leq r_p \leq 9.6\mpch$)
\label{tab9}}
\tablewidth{0pt} \tablehead{
    Sample & ${\rm F1_{blue}}$ & ${\rm F2_{blue}}$ & ${\rm F3_{blue}}$ & ${\rm
    F4_{blue}}$ & ${\rm F5_{blue}}$ & ${\rm F6_{blue}}$ }

\startdata
& $\alpha$ & $\alpha$  & $\alpha$  & $\alpha$  & $\alpha$ & $\alpha$   \\
& $r_0$    & $r_0$     & $r_0$     & $r_0$     & $r_0$    & $r_0$   \\
\tableline

${\rm F1_{blue}}$ & $ 1.89 \pm 0.12$ & & & & &\\
     & $ 6.45 \pm 1.03$ & & & & &\\

${\rm F2_{blue}}$ & $ 1.75 \pm 0.07$ &$ 1.75 \pm 0.06$ & & & &\\
     & $ 5.50 \pm 0.69$ &$ 5.11 \pm 0.33$ & & & &\\

${\rm F3_{blue}}$ & $ 1.83 \pm 0.06$ &$ 1.73 \pm 0.04$ &$ 1.74 \pm 0.04$ & & &\\
     & $ 5.17 \pm 0.36$ &$ 4.71 \pm 0.27$ &$ 4.63 \pm 0.46$ & & &\\

${\rm F4_{blue}}$ & $ 1.82 \pm 0.10$ &$ 1.72 \pm 0.09$ &$ 1.72 \pm 0.06$ &$ 1.69 \pm 0.09$ & &\\
     & $ 4.97 \pm 0.45$ &$ 4.62 \pm 0.21$ &$ 4.39 \pm 0.32$ &$ 4.14 \pm 0.43$ & &\\

${\rm F5_{blue}}$ & $ 1.78 \pm 0.18$ &$ 1.65 \pm 0.13$ &$ 1.69 \pm 0.09$ &$ 1.67 \pm 0.12$ &$ 1.67 \pm 0.15$ &\\
     & $ 4.85 \pm 0.41$ &$ 4.29 \pm 0.46$ &$ 4.13 \pm 0.27$ &$ 3.90 \pm 0.38$ &$ 3.71 \pm 0.44$ &\\

${\rm F6_{blue}}$ & &$ 1.64 \pm 0.04$ &$ 1.78 \pm 0.04$ &$ 1.77 \pm 0.04$ &$ 1.73 \pm 0.04$ &$ 1.74 \pm 0.09$\\
     & &$ 4.71 \pm 1.49$ &$ 4.10 \pm 0.57$ &$ 3.93 \pm 0.35$ &$ 3.61 \pm 0.34$ &$ 3.46 \pm 0.33$\\

\tableline \tableline

$\alpha=\overline{\alpha}=1.74$ & $r_0$ & $r_0$ & $r_0$ & $r_0$ & $r_0$ & $r_0$  \\

 \tableline

${\rm F1_{blue}}$ & $6.15 \pm 0.58$ & & & & & \\

${\rm F2_{blue}}$ & $5.47 \pm 0.23$ &$5.08 \pm 0.26$ & & & & \\

${\rm F3_{blue}}$ & $4.99 \pm 0.17$ &$4.75 \pm 0.21$ &$4.62 \pm 0.24$ & & & \\

${\rm F4_{blue}}$ & $4.81 \pm 0.21$ &$4.66 \pm 0.31$ &$4.42 \pm 0.24$ &$4.22 \pm 0.23$ & & \\

${\rm F5_{blue}}$ & $4.77 \pm 0.36$ &$4.47 \pm 0.29$ &$4.21 \pm 0.21$ &$4.03 \pm 0.18$ &$3.80 \pm 0.13$ & \\

${\rm F6_{blue}}$ & &$4.90 \pm 0.69$ &$4.02 \pm 0.23$ &$3.89 \pm 0.37$
&$3.62 \pm 0.19$ &$3.46 \pm 0.13$

\enddata
\end{deluxetable*}

\begin{deluxetable*}{lcccccc}
  \tablecaption{The correlation length and slope for flux-limited
    samples of blue galaxies on small scales
    ($0.16\mpch \leq r_p \leq 0.98\mpch$)
\label{tab10}} \tablewidth{0pt}
  \tablehead{ Sample & ${\rm F1_{blue}}$ & ${\rm F2_{blue}}$ & ${\rm
  F3_{blue}}$ & ${\rm F4_{blue}}$ & ${\rm F5_{blue}}$ & ${\rm F6_{blue}}$ }

\startdata
& $\alpha$ & $\alpha$  & $\alpha$  & $\alpha$  & $\alpha$ & $\alpha$   \\
& $r_0$    & $r_0$     & $r_0$     & $r_0$     & $r_0$    & $r_0$   \\
\tableline

${\rm F1_{blue}}$ & $ 1.31 \pm 0.41$ & & & & & \\
    & $ 8.52 \pm 2.25$ & & & & & \\

${\rm F2_{blue}}$ & $ 2.07 \pm 0.11$ &$ 1.70 \pm 0.06$ & & & & \\
    & $ 4.72 \pm 0.55$ &$ 5.21 \pm 0.55$ & & & & \\

${\rm F3_{blue}}$ & $ 1.63 \pm 0.12$ &$ 1.80 \pm 0.06$ &$ 1.64 \pm 0.04$ & & & \\
    & $ 5.45 \pm 0.78$ &$ 4.44 \pm 0.21$ &$ 4.65 \pm 0.22$ & & & \\

${\rm F4_{blue}}$ & $ 1.69 \pm 0.15$ &$ 1.81 \pm 0.11$ &$ 1.64 \pm 0.06$ &$ 1.65 \pm 0.06$ & & \\
    & $ 5.30 \pm 0.45$ &$ 4.14 \pm 0.43$ &$ 4.21 \pm 0.23$ &$ 4.06 \pm 0.38$ & & \\

${\rm F5_{blue}}$ & $ 1.99 \pm 0.33$ &$ 1.81 \pm 0.14$ &$ 1.64 \pm 0.07$ &$ 1.61 \pm 0.08$ &$ 1.62 \pm 0.07$ & \\
    & $ 4.11 \pm 0.88$ &$ 3.86 \pm 0.49$ &$ 4.03 \pm 0.41$ &$ 3.88 \pm 0.38$ &$ 3.60 \pm 0.39$ & \\

${\rm F6_{blue}}$ & &$ 1.62 \pm 0.26$ &$ 1.73 \pm 0.32$ &$ 1.71 \pm 0.21$ &$ 1.66 \pm 0.30$ &$ 1.60 \pm 0.28$ \\
    & &$ 5.10 \pm 1.92$ &$ 3.98 \pm 0.97$ &$ 3.91 \pm 0.52$ &$ 3.47 \pm 0.42$ &$ 3.45 \pm 0.51$ \\

\tableline \tableline
$\alpha=\overline{\alpha}=1.70$ & $r_0$ & $r_0$ & $r_0$ & $r_0$ & $r_0$ & $r_0$  \\
 \tableline

${\rm F1_{blue}}$ & $6.13 \pm 0.96$ & & & & & \\

${\rm F2_{blue}}$ & $5.96 \pm 0.26$ &$5.22 \pm 0.25$ & & & & \\

${\rm F3_{blue}}$ & $5.23 \pm 0.38$ &$4.71 \pm 0.27$ &$4.42 \pm 0.28$ & & & \\

${\rm F4_{blue}}$ & $5.33 \pm 0.35$ &$4.40 \pm 0.22$ &$4.09 \pm 0.16$ &$3.96 \pm 0.22$ & & \\

${\rm F5_{blue}}$ & $4.77 \pm 0.35$ &$4.10 \pm 0.27$ &$3.92 \pm 0.24$ &$3.73 \pm 0.23$ &$3.53 \pm 0.15$ & \\

${\rm F6_{blue}}$ & &$4.89 \pm 0.66$ &$4.05 \pm 0.22$ &$3.94 \pm 0.17$
&$3.44 \pm 0.17$ &$3.33 \pm 0.13$

\enddata

\end{deluxetable*}

\begin{deluxetable*}{lcccccc}
  \tablecaption{The correlation length and slope
   between red and blue galaxies
   on large scales ($0.98\mpch \leq r_p \leq 9.6\mpch$)
    \label{tab11}}
  \tablewidth{0pt}
  \tablehead { Sample & ${\rm F1_{red}}$ & ${\rm F2_{red}}$ & ${\rm F3_{red}}$
  & ${\rm F4_{red}}$ & ${\rm F5_{red}}$ & ${\rm F6_{red}}$ }

\startdata
 & $\alpha$ & $\alpha$  & $\alpha$  & $\alpha$  & $\alpha$ & $\alpha$   \\
 & $r_0$    & $r_0$     & $r_0$     & $r_0$     & $r_0$    & $r_0$   \\
\tableline

${\rm F1_{blue}}$& $ 1.79 \pm 0.05$ &$ 1.81 \pm 0.04$ &$ 1.80 \pm 0.04$ &$ 1.89 \pm 0.04$ &$ 1.91 \pm 0.10$ &\\
   & $ 6.80 \pm 0.94$ &$ 6.10 \pm 0.46$ &$ 5.99 \pm 0.41$ &$ 6.16 \pm 0.55$ &$ 6.16 \pm 1.44$ &\\

${\rm F2_{blue}}$& $ 1.74 \pm 0.03$ &$ 1.71 \pm 0.02$ &$ 1.76 \pm 0.02$ &$ 1.73 \pm 0.03$ &$ 1.71 \pm 0.06$ &$ 1.64 \pm 0.13$\\
   & $ 6.20 \pm 0.41$ &$ 5.61 \pm 0.18$ &$ 5.45 \pm 0.19$ &$ 5.48 \pm 0.27$ &$ 5.34 \pm 0.45$ &$ 6.01 \pm 1.76$\\

${\rm F3_{blue}}$& $1.79 \pm 0.03$ &$1.73 \pm 0.02$ &$1.73 \pm 0.03$ &$1.74 \pm 0.03$ &$1.75 \pm 0.03$ &$1.82 \pm 0.14$\\
   & $6.03 \pm 0.36$ &$5.49 \pm 0.24$ &$5.21 \pm 0.22$ &$5.30 \pm 0.35$ &$5.18 \pm 0.26$ &$5.43 \pm 0.51$\\

${\rm F4_{blue}}$& $1.74 \pm 0.06$ &$1.76 \pm 0.04$ &$1.74 \pm 0.03$ &$1.71 \pm 0.04$ &$1.74 \pm 0.03$ &$1.85 \pm 0.11$\\
   & $ 5.67 \pm 0.63$ &$ 5.33 \pm 0.24$ &$ 5.04 \pm 0.27$ &$ 5.05 \pm 0.25$ &$ 5.00 \pm 0.21$ &$ 5.47 \pm 0.36$\\

${\rm F5_{blue}}$& $ 1.72 \pm 0.10$ &$ 1.71 \pm 0.05$ &$ 1.68 \pm 0.04$ &$ 1.70 \pm 0.03$ &$ 1.73 \pm 0.02$ &$ 1.79 \pm 0.06$\\
   & $ 5.44 \pm 0.72$ &$ 5.06 \pm 0.27$ &$ 4.85 \pm 0.24$ &$ 4.87 \pm 0.22$ &$ 4.79 \pm 0.18$ &$ 5.00 \pm 0.14$\\

${\rm F6_{blue}}$& &$ 1.75 \pm 0.20$ &$ 1.74 \pm 0.12$ &$ 1.75 \pm 0.09$ &$ 1.83 \pm 0.06$ &$ 1.86 \pm 0.08$ \\
   & &$ 5.27 \pm 1.35$ &$ 4.90 \pm 0.44$ &$ 4.68 \pm 0.24$ &$ 4.71 \pm 0.23$ &$ 4.83 \pm 0.14$ \\

\tableline \tableline
 $\alpha=\overline{\alpha}=1.76$ & $r_0$ & $r_0$ & $r_0$ & $r_0$ & $r_0$ & $r_0$  \\
 \tableline

${\rm F1_{blue}}$ & $6.74 \pm 0.53$ &$6.02 \pm 0.15$ &$5.92 \pm 0.18$ &$6.00 \pm 0.15$ &$6.00 \pm 0.21$ &\\

${\rm F2_{blue}}$ & $6.23 \pm 0.14$ &$5.69 \pm 0.11$ &$5.44 \pm 0.19$ &$5.52 \pm 0.16$ &$5.39 \pm 0.15$ &$6.17 \pm 0.58$ \\

${\rm F3_{blue}}$ & $5.98 \pm 0.12$ &$5.52 \pm 0.09$ &$5.25 \pm 0.12$ &$5.32 \pm 0.15$ &$5.20 \pm 0.15$ &$5.34 \pm 0.14$ \\

${\rm F4_{blue}}$ & $5.70 \pm 0.16$ &$5.32 \pm 0.22$ &$5.06 \pm 0.18$ &$5.10 \pm 0.17$ &$5.03 \pm 0.18$ &$5.33 \pm 0.15$ \\

${\rm F5_{blue}}$ & $5.51 \pm 0.11$ &$5.14 \pm 0.11$ &$4.94 \pm 0.17$ &$4.93 \pm 0.13$ &$4.82 \pm 0.09$ &$4.97 \pm 0.16$ \\

${\rm F6_{blue}}$ & &$5.29 \pm 0.73$ &$4.92 \pm 0.27$ &$4.69 \pm 0.35$
&$4.62 \pm 0.19$ &$4.68 \pm 0.17$

\enddata
\end{deluxetable*}

\begin{deluxetable*}{lcccccc}
  \tablecaption{The correlation length and slope
   between red and blue galaxies
   on small scales ($0.16\mpch \leq r_p \leq 0.98\mpch$)
\label{tab12}}
  \tablewidth{0pt} \tablehead { Sample & ${\rm F1_{red}}$ & ${\rm F2_{red}}$ &
    ${\rm F3_{red}}$ & ${\rm F4_{red}}$ & ${\rm F5_{red}}$
    & ${\rm F6_{red}}$}

\startdata
 & $\alpha$ & $\alpha$  & $\alpha$  & $\alpha$  & $\alpha$ & $\alpha$   \\
 & $r_0$    & $r_0$     & $r_0$     & $r_0$     & $r_0$    & $r_0$   \\
\tableline

${\rm F1_{blue}}$&$ 1.87 \pm 0.16$ &$ 2.01 \pm 0.13$ &$ 2.06 \pm 0.09$ &$ 1.84 \pm 0.14$ &$ 1.98 \pm 0.15$ &\\
   &$ 7.60 \pm 0.96$ &$ 5.08 \pm 0.43$ &$ 4.90 \pm 0.41$ &$ 6.14 \pm 0.56$ &$ 5.74 \pm 0.53$ &\\

${\rm F2_{blue}}$&$ 2.04 \pm 0.17$ &$ 1.81 \pm 0.13$ &$ 1.74 \pm 0.08$ &$ 1.74 \pm 0.10$ &$ 1.67 \pm 0.16$ &$ 1.59 \pm 0.16$\\
   &$ 5.15 \pm 0.35$ &$ 5.31 \pm 0.35$ &$ 5.34 \pm 0.44$ &$ 5.50 \pm 0.36$ &$ 5.39 \pm 0.44$ &$ 6.86 \pm 1.89$\\

${\rm F3_{blue}}$&$1.77 \pm 0.16$ &$1.77 \pm 0.12$ &$1.67 \pm 0.15$ &$1.67 \pm 0.07$ &$1.65 \pm 0.06$ &$1.69 \pm 0.12$\\
   &$6.03 \pm 0.53$ &$5.19 \pm 0.44$ &$5.27 \pm 0.34$ &$5.23 \pm 0.48$ &$5.28 \pm 0.26$ &$6.00 \pm 0.59$\\

${\rm F4_{blue}}$&$1.92 \pm 0.19$ &$1.68 \pm 0.15$ &$1.67 \pm 0.14$ &$1.67 \pm 0.11$ &$1.64 \pm 0.08$ &$1.71 \pm 0.12$\\
   &$ 5.12 \pm 0.45$ &$ 5.36 \pm 0.56$ &$ 5.11 \pm 0.68$ &$ 5.11 \pm 0.46$ &$ 5.04 \pm 0.37$ &$ 5.82 \pm 0.45$\\

${\rm F5_{blue}}$&$ 1.85 \pm 0.12$ &$ 1.84 \pm 0.16$ &$ 1.73 \pm 0.08$ &$ 1.65 \pm 0.13$ &$ 1.56 \pm 0.06$ &$ 1.67 \pm 0.11$\\
   &$ 5.28 \pm 0.71$ &$ 4.71 \pm 0.58$ &$ 4.71 \pm 0.42$ &$ 4.89 \pm 0.45$ &$ 5.09 \pm 0.46$ &$ 5.16 \pm 0.48$\\

${\rm F6_{blue}}$& &$ 2.07 \pm 0.27$ &$ 1.79 \pm 0.24$ &$ 1.73 \pm 0.23$ &$ 1.68 \pm 0.11$ &$ 1.70 \pm 0.12$ \\
   & &$ 4.29 \pm 1.47$ &$ 4.92 \pm 0.48$ &$ 4.66 \pm 0.27$ &$ 4.77 \pm 0.51$ &$ 5.04 \pm 0.49$ \\

\tableline \tableline
 $\alpha=\overline{\alpha}=1.76$ & $r_0$ & $r_0$ & $r_0$ & $r_0$ & $r_0$ & $r_0$  \\
 \tableline

${\rm F1_{blue}}$ & $7.22 \pm 0.61$ &$5.94 \pm 0.28$ &$5.98 \pm 0.31$ &$6.53 \pm 0.36$ &$6.77 \pm 0.32$ & \\

${\rm F2_{blue}}$ & $6.26 \pm 0.31$ &$5.52 \pm 0.16$ &$5.25 \pm 0.13$ &$5.43 \pm 0.16$ &$5.20 \pm 0.21$ &$6.02 \pm 0.55$ \\

${\rm F3_{blue}}$ & $6.06 \pm 0.37$ &$5.22 \pm 0.13$ &$4.93 \pm 0.12$ &$5.02 \pm 0.15$ &$4.97 \pm 0.15$ &$5.74 \pm 0.26$ \\

${\rm F4_{blue}}$ & $5.69 \pm 0.41$ &$5.11 \pm 0.35$ &$4.80 \pm 0.34$ &$4.79 \pm 0.38$ &$4.75 \pm 0.35$ &$5.61 \pm 0.28$ \\

${\rm F5_{blue}}$ & $5.59 \pm 0.27$ &$4.98 \pm 0.33$ &$4.61 \pm 0.25$ &$4.53 \pm 0.34$ &$4.44 \pm 0.19$ &$4.91 \pm 0.27$ \\

${\rm F6_{blue}}$ & &$5.20 \pm 0.75$ &$5.01 \pm 0.24$ &$4.57 \pm 0.19$
&$4.58 \pm 0.28$ &$4.90 \pm 0.24$

\enddata
\end{deluxetable*}

\begin{figure}
  \plotone{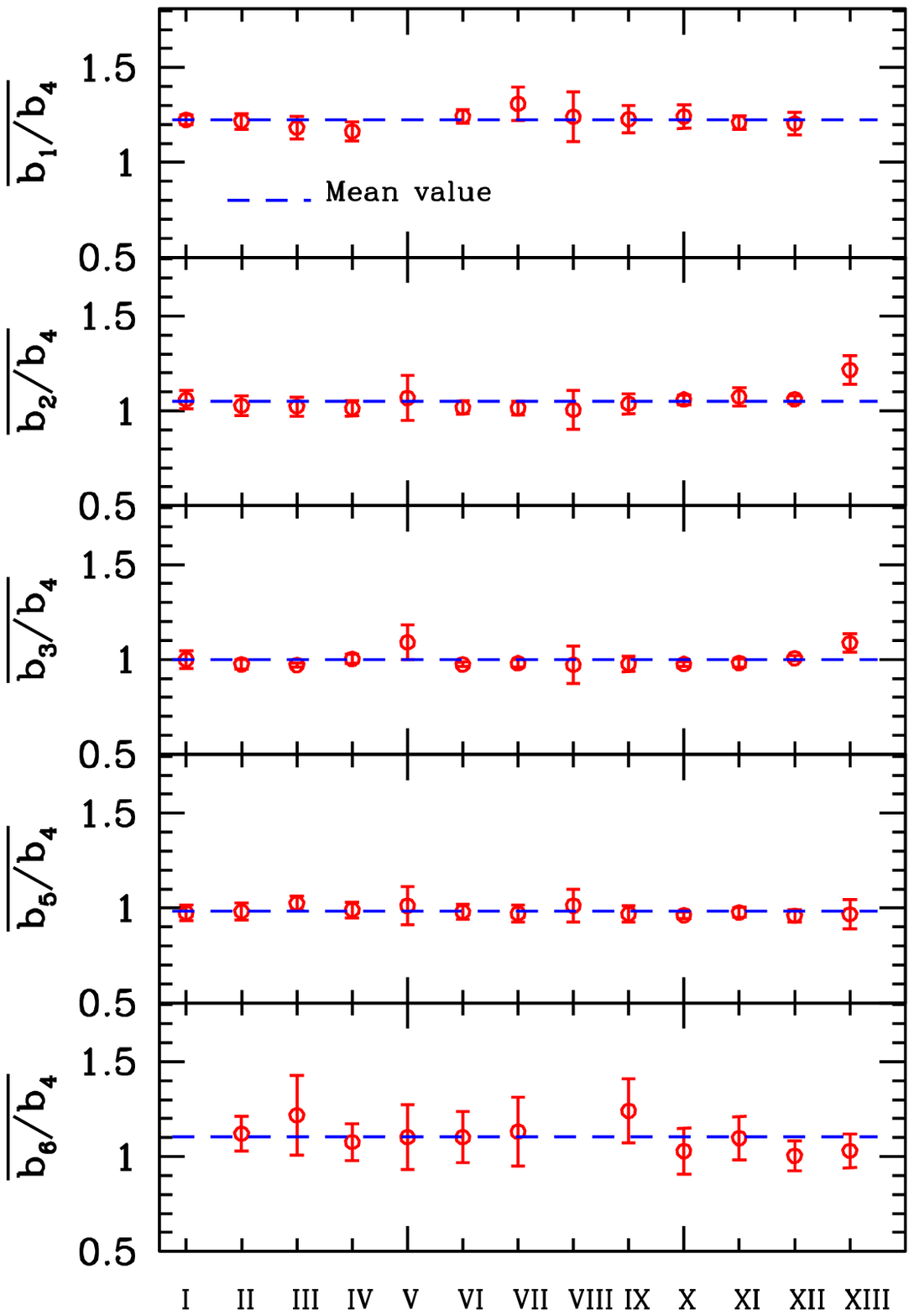}  \caption{The  relative  bias parameters  for  red
    galaxies obtained from  the correlation functions over $0.98\mpch$
    to $9.8\mpch$ using 13 different approaches (see the main text for
    details).  The  error-bars are  based on the  scatter of  the data
    points in the range $0.98\mpch< r_p\le 9.8\mpch$.  Panels from top
    to bottom  correspond to the bias parameters  for ${\rm Fi_{red}}$
    ($i=1,2,3,5,6$),  normalized  by  that  of ${\rm  F4_{red}}$.  The
    horizontal axis  labels the different approaches that  are used to
    determine  the  bias parameter  for  each  subsamples.  Some  data
    points are missing, because the redshift overlap between F1 and F6
    is   too  small  to   measure  their   cross-correlation  function
    reliably.}
\label{biasR2}
\end{figure}

\begin{figure}
  \plotone{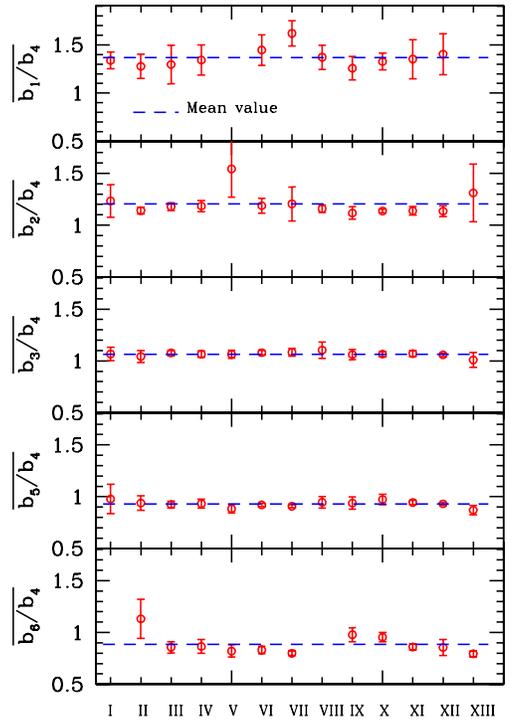}  \caption{The same  as Fig.~\ref{biasR2}  for blue
    galaxies.   Panels  from top  to  bottom  correspond  to the  bias
    parameters  for ${\rm  Fi_{blue}}$ ($i=1,2,3,5,6$),  normalized by
    that of ${\rm F4_{blue}}$.}
\label{biasB2}
\end{figure}

\begin{figure*}
  \plotone{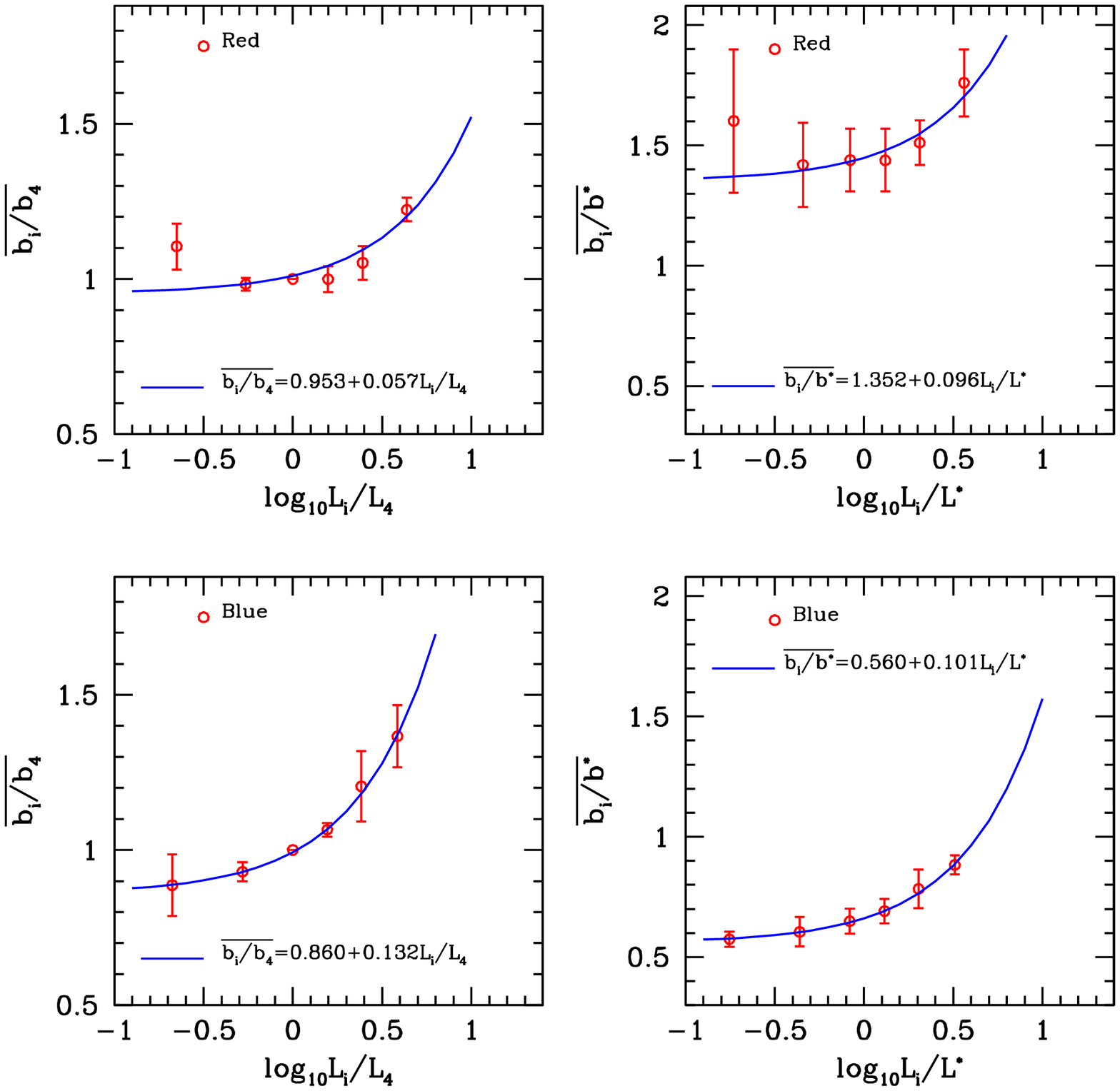}  \caption{Left panels:  The average  relative bias
    parameter,    obtained    using    the   correlation    data    in
    $0.98\mpch<r_p\le  9.8\mpch$ as a  function of  galaxy luminosity,
    for red (upper panel) and blue (lower panel) galaxies. The results
    are  obtained using the  flux-limited samples,  and both  the bias
    parameters and  the luminosities are  normalized by the  values of
    ${\rm F4_{red}}$ and ${\rm  F4_{blue}}$, respectively. For a given
    luminosity,  the mean  and error-bar  are based  on  the different
    measurements  shown in  Figs.~\ref{biasR2}  and \ref{biasB2}.  The
    solid curves show fits to the data.  Right panels: The same as the
    left  panels,  except  that  here  the  bias  parameters  and  the
    luminosities  are normalized by  the values  for $L^*$  galaxies.}
  \label{biasrb}
\end{figure*}

In  order  to study  how  galaxy  clustering  depends on  both  galaxy
luminosity and color, we have subdivided each flux-limited sample, Fi,
into a red and a blue subsample according to the criteria described in
equation (\ref{color_criterion}). We  remind that these subsamples are
referred to as ${\rm Fi_{red}}$ and ${\rm Fi_{blue}}$. We have carried
out the  same analyses as described  in Section~\ref{sec:result_L} for
these color  samples.  Figs.~\ref{fig7} shows  the correlation lengths
$r_0$  and  slopes  $\alpha$   obtained  by  fitting  the  correlation
functions  in  the  range  $0.98\mpch  \leq r_p  \leq  9.6\mpch$;  the
corresponding  results  for small  scales,  $0.16\mpch  \leq r_p  \leq
0.98\mpch$, are  shown in Figs.~\ref{fig8}.  In both figures,  we also
include  the  results  for  the  total  samples  (i.e.  without  color
separation) for  comparison.  Again, the error-bars  are obtained from
the scatter among the three  regions in the SDSS observations. To make
the results accessible to others, we  list all the values of $r_0$ and
$\alpha$ in Tables~\ref{tab7} (red galaxies; large scales), \ref{tab8}
(red  galaxies;  small   scales),  \ref{tab9}  (blue  galaxies;  large
scales), and \ref{tab10} (blue  galaxies; small scales).  In all these
tables,  we also  include  the best  fit  values for  $r_0$ by  fixing
$\alpha={\overline\alpha}$.

As one  can see  clearly from the  left panels in  Fig.~\ref{fig7}, on
large  scales red galaxies  in all  cases have  larger $r_0$  than all
galaxies and blue galaxies in  the same luminosity bin. Note also that
the  luminosity dependence  of  $r_0$ is  seen  in all,  red and  blue
samples, although the dependence is  stronger for blue galaxies, as to
be  quantified  in  the  following. Overall,  the  correlation  slope,
$\alpha$ (shown in the right  panel of Fig.~\ref{fig7}), is larger for
red galaxies,  and the dependence is  stronger for the  cases where at
least one of the two samples in cross-correlation is faint.

As shown  in Fig.~\ref{fig8},  the correlation lengths  $r_0$ obtained
from the  projected 2PCFs on  small scales have properties  similar to
those on large scales, in  that red galaxies have a larger correlation
length.  But the  trend is  quite different  in detail.   Firstly, the
difference  between the  correlation lengths  obtained from  the cross
correlations  of the brightest  red sample  ($\FIred$) with  other red
samples   are   significantly  larger   here   than   that  shown   in
Fig.~\ref{fig7}.   Since  the   brightest  red  galaxies  are  located
predominately in clusters and rich  groups, this suggests that many of
 the faint red galaxies  are also in clusters and groups. Secondly,
for $\FIIIred$ through $\FVred$, their cross-correlation with the {\it
  faintest} red galaxies has an amplitude that is significantly larger
than their cross-correlation with  brighter (except the brightest) red
galaxies.  This  indicates  that  many red  galaxies  of  intermediate
luminosities may possess halos of red satellite galaxies that are more
than   1   magnitudes  fainter.   Finally,   on   small  scales,   the
luminosity-dependence $r_0$ for blue galaxies appears to be weaker.

It is also interesting to  look at the cross correlation {\it between}
red  and  blue  galaxies.   In Figs.~\ref{fig9}  and  \ref{fig10},  we
compare the values  of $r_0$ and $\alpha$ for  such cross correlations
on large  and small scales with  those for all  galaxies. These values
are also listed in Tables \ref{tab11} and \ref{tab12}. As one can see,
on    large   scales    ($0.98\mpch   \leq    r_p    \leq   9.6\mpch$)
(Fig.~\ref{fig9}), the  values of $r_0$ and $\alpha$  for the red/blue
cross-correlation  functions  are  quite  similar  to  those  for  all
galaxies.  However,  there is  a  marked  difference  on small  scales
($0.16\mpch  \leq r_p  \leq 0.98\mpch$)  (Fig.~\ref{fig10}).  Here the
cross  correlation  amplitudes  between  the  brightest  red  galaxies
($\FIred$) and  blue galaxies are  significantly lower than  those for
all  galaxies.  Since many  of  the  brightest  red galaxies  are  the
brightest  central  galaxies  in  clusters  and  groups,  this  result
suggests the  satellite galaxies of these  galaxies are preferentially
red, in  qualitative agreement  with the findings  by Weinmann et  al.
(2006).

In order to quantify how galaxies of different colors and luminosities
are biased with respect to each  other on large scales, we measure the
relative bias factors for red and blue galaxies based on the ratios of
the    corresponding     projected    2PCFs,    as     we    did    in
Section~\ref{sec:result_L} for  the luminosity samples.  Note that, in
addition  to  the red-red  and  blue-blue cross-correlation  functions
among  the  6   luminosity  bins,  one  can  also   use  the  red-blue
cross-correlation functions to obtain the relative bias. These are
\begin{equation}\label{eq:bi_b4_R}
\frac{b_i}{b_4}=\frac{W_{Ri,Bj}}{W_{R4,Bj}}\,
\end{equation}
for red galaxies, and
\begin{equation}\label{eq:bi_b4_B}
\frac{b_i}{b_4}=\frac{W_{Bi,Rj}}{W_{B4,Rj}}\,
\end{equation}
for blue galaxies, where $i=1,2,3,5,6$ and $j=1,2,3,4,5,6$, and Bi and
Ri denotes  ${\rm Fi_{blue}}$ and ${\rm  Fi_{red}}$, respectively.  In
what  follows, these  six approaches  will be  referred  as approaches
VIII,     IX,     $\cdot\cdot\cdot$,     XIII,    corresponding     to
$j=1,2,\cdot\cdot\cdot, 6$,  respectively. Together with  the previous
seven approaches,  we have now in  total 13 approaches  to measure the
relative   bias  $b_i/b_4$   for  each   red  or   blue   subsamples.
Figs.~\ref{biasR2} and \ref{biasB2} show the large-scale relative bias
factors obtained  with the  13 approaches for  red and  blue galaxies,
respectively. Some  points are missing  again because the  overlaps of
the samples in  question are too small. The dashed  line in each panel
is the average value of the relative bias. As one can see, most of the
approaches give  extremely consistent results. The  biggest outlier is
the one that corresponds to the cross correlation $\FIIred$-$\FVIblue$
(approach XIII in the  second row of Fig~\ref{biasR2}), which deviates
from the mean bias ratio by just over $2\sigma$.

In the left panels of  Fig.~\ref{biasrb} (upper panel for red galaxies
and lower panel for blue  galaxies), we show the average relative bias
($\overline{b_i/b_4}$   obtained  from   the  13   approaches)  versus
luminosity $L_i/L_4$, where $L_i$ for  each sample is estimated in the
same way  as described in Section \ref{sec:result_L}.   Fitting to the
data with  a linear  model gives $\overline{b_i/b_4}  = 0.953  + 0.057
L_i/L_4$  for red galaxies,  and $\overline{b_i/b_4}  = 0.860  + 0.132
L_i/L_4$ for blue galaxies.  As one can see, the luminosity-dependence
is much  stronger for  blue galaxies than  for red galaxies.   For red
galaxies,  significant  luminosity-dependence  is  seen  only  at  the
brightest end,  and there is  indication that the bias  factor becomes
larger for the faintest sample.

In order  to put the bias  factors for different galaxies  in the same
absolute scale, we normalize all bias factors relative to that for all
the galaxies (blue and red) in the F* sample, i.e. we define
\begin{equation}
\overline{b_i/b^*}
=\overline{b_i/b_4} \sqrt{\frac{W_{R4,R4}}{W^{*,*}}}
\end{equation}
for red galaxies, and
\begin{equation}
\overline{b_i/b^*}=\overline{b_i/b_4} \sqrt{\frac{W_{B4,B4}}{W^{*,*}}}
\end{equation}
for blue  galaxies.  Here  $W_{R4,R4}$, $W_{B4,B4}$ and  $W^{*,*}$ are
the  projected   auto  2PCFs  for  $F4_{red}$,   $F4_{blue}$  and  F*,
respectively.  The  values of $\overline{b_i/b^*}$  versus $L/L^*$ are
shown in  the right panels  of Fig.~\ref{biasrb} (upper panel  for red
galaxies and lower  panel for blue galaxies). Normalized  in this way,
the bias factors  for red galaxies are all larger  than 1, while those
for  blue galaxies  smaller than  1, for  the luminosity  range probed
here. Fitting  the luminosity dependence  with a linear model,  we get
$\overline{b_i/b^*}  = 1.352 +  0.096 L_i/L^*$  for red  galaxies, and
$\overline{b_i/b^*} =  0.560 + 0.101 L_i/L^*$ for  blue galaxies. Note
that the  constant component  in the relation  is much larger  for red
galaxies than  for blue  galaxies, although the  linear terms  are now
similar. These  results are in  good agreement with those  obtained in
Zehavi et al.  (2005) based on auto-correlation functions.

\section{Non-linear and stochastic bias}
\label{sec:discussion}

\begin{figure*}
  \plotone{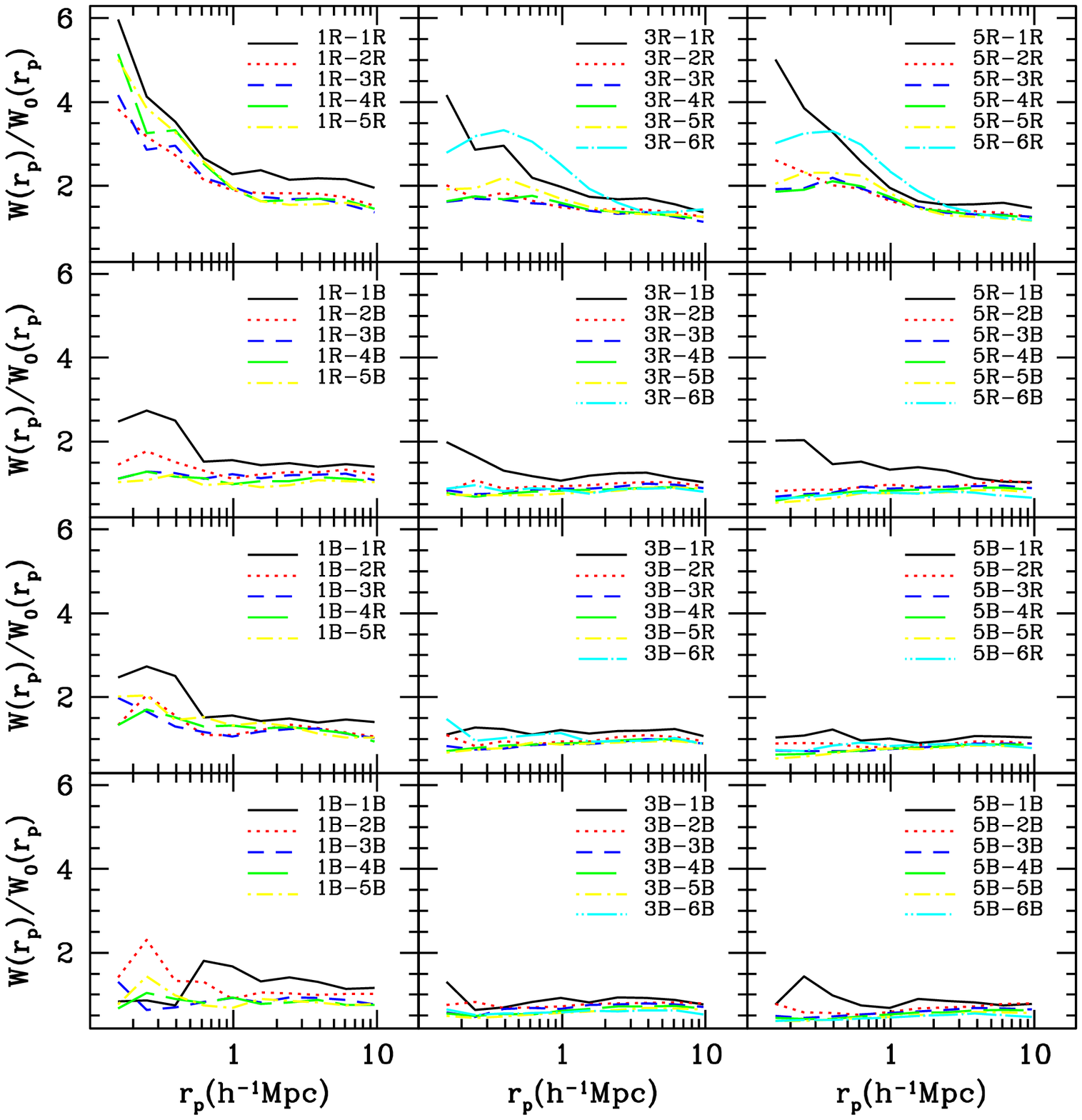}   \caption{The   ratios   between  the   projected
    correlation  function  $W(r_p)$  and  $W_0(r_p)$  for  cases  with
    various combinations of color  and luminosity, where $W_0(r_p)$ is
    the power-law fit to  the projected correlation function of sample
    $F^*$ on large scales ($r_0=5.49\mpch$, $\alpha=1.76$).  Three top
    panels show the  results for red galaxies, six  middle panels show
    the results between red and blue galaxies, and three bottom panels
    show the results  for blue galaxies. In each  panel, iR-jR denotes
    ${\rm  Fi_{red}}$  with  ${\rm  Fj_{red}}$,  iB-jB  denotes  ${\rm
      Fi_{blue}}$   with  ${\rm   Fj_{blue}}$,  iR-jB   denotes  ${\rm
      Fi_{red}}$ with ${\rm Fj_{blue}}$.}
\label{wrp}
\end{figure*}

\begin{figure*}
  \plotone{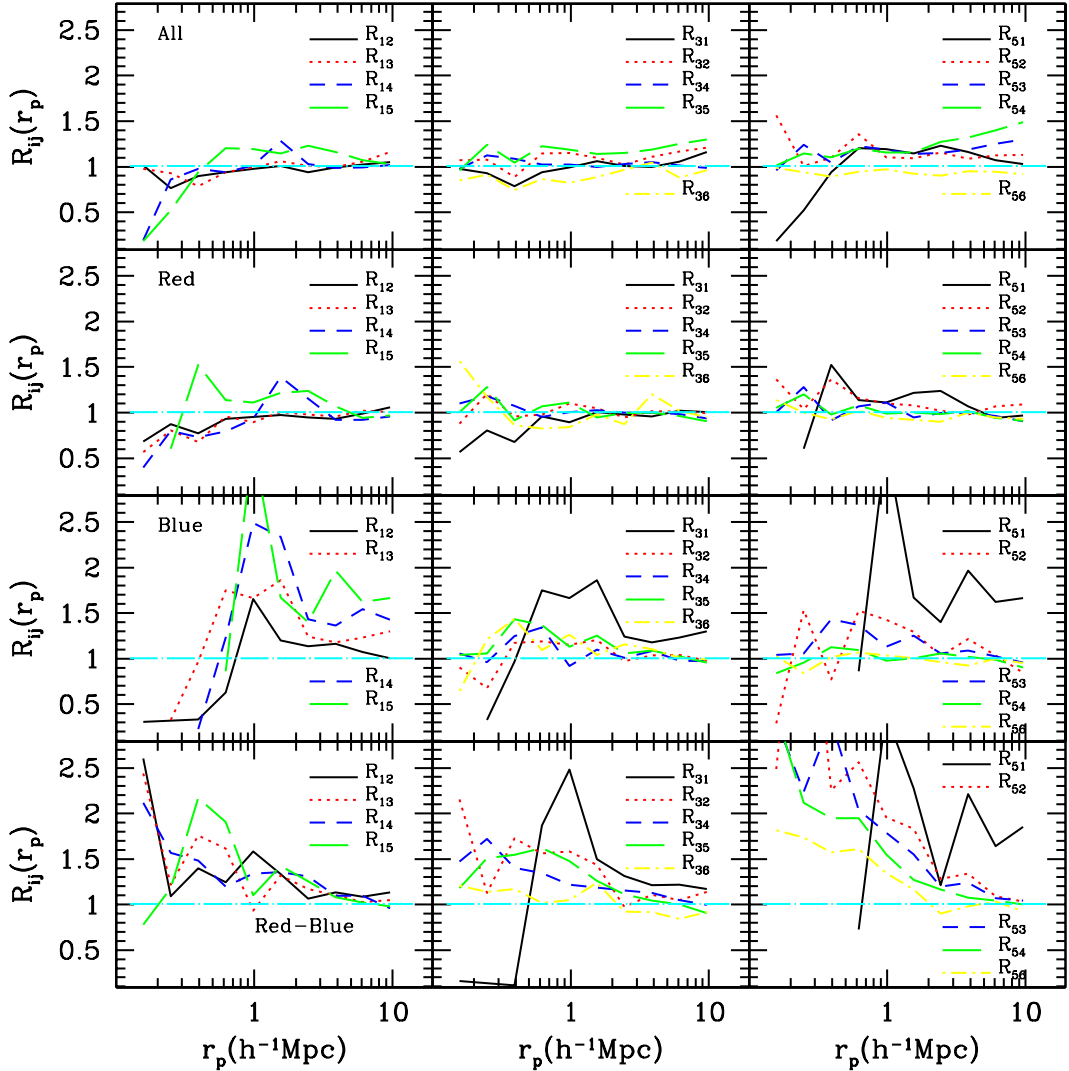}  \caption{The  value   of  ${\cal  R}_{ij}$  as  a
    function of $r_p$ for cases with various combinations of color and
    luminosity.  Panels  from top to bottom show  results for all-all,
    red-red,  blue-blue,  and  red-blue  samples,  respectively.   The
    straight line in each panel shows ${\cal R}_{ij}=1$.}
\label{fig_Rij}
\end{figure*}

\begin{figure*}
  \plotone{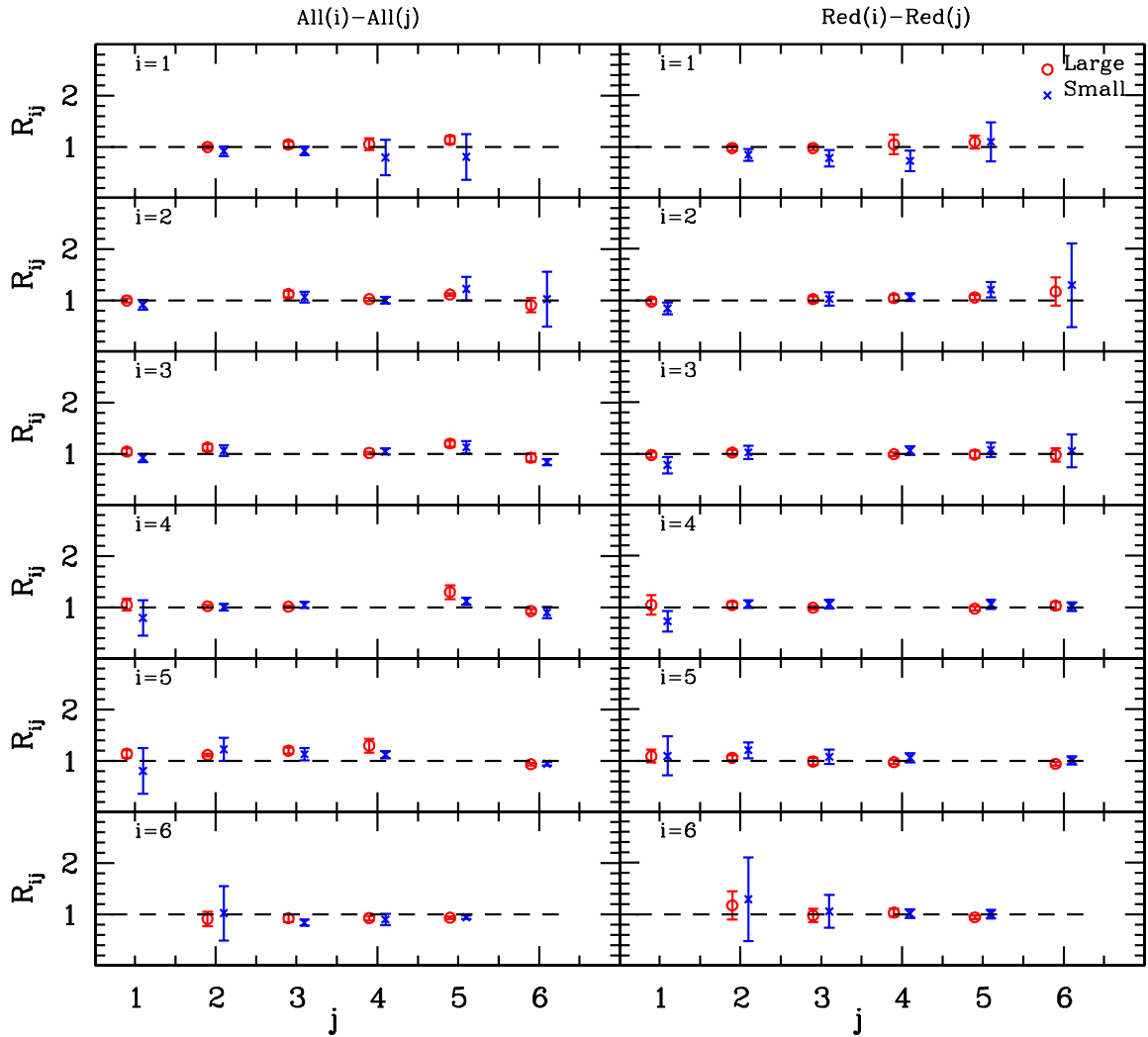}  \caption{The  ratio  ${\cal R}_{ij}$  defined  in
    equation (\ref{ratio:bias})  for all (left panels)  and red (right
    panels)  galaxies. In all  panels, circles  show results  based on
    data on  large scales, $0.98\mpch  \leq r_p \leq  9.6\mpch$, while
    crosses show  results based data on small  scales, $0.16\mpch \leq
    r_p  \leq  0.98\mpch$.   From   top  to  bottom  are  results  for
    $i=1,2,3,4,5,6$, respectively.  The value of $j$ for  each case is
    labeled  on the  horizontal axis.  The dashed  line in  each panel
    shows ${\cal  R}_{ij}=1$. The  error-bars are estimated  using the
    scatter of  the data  points on large  scales $0.98\mpch  \leq r_p
    \leq  9.6\mpch$   and  small  scales  $0.16\mpch   \leq  r_p  \leq
    0.98\mpch$, respectively.}
\label{fig:stochasticity1}
\end{figure*}
\begin{figure*}
  \plotone{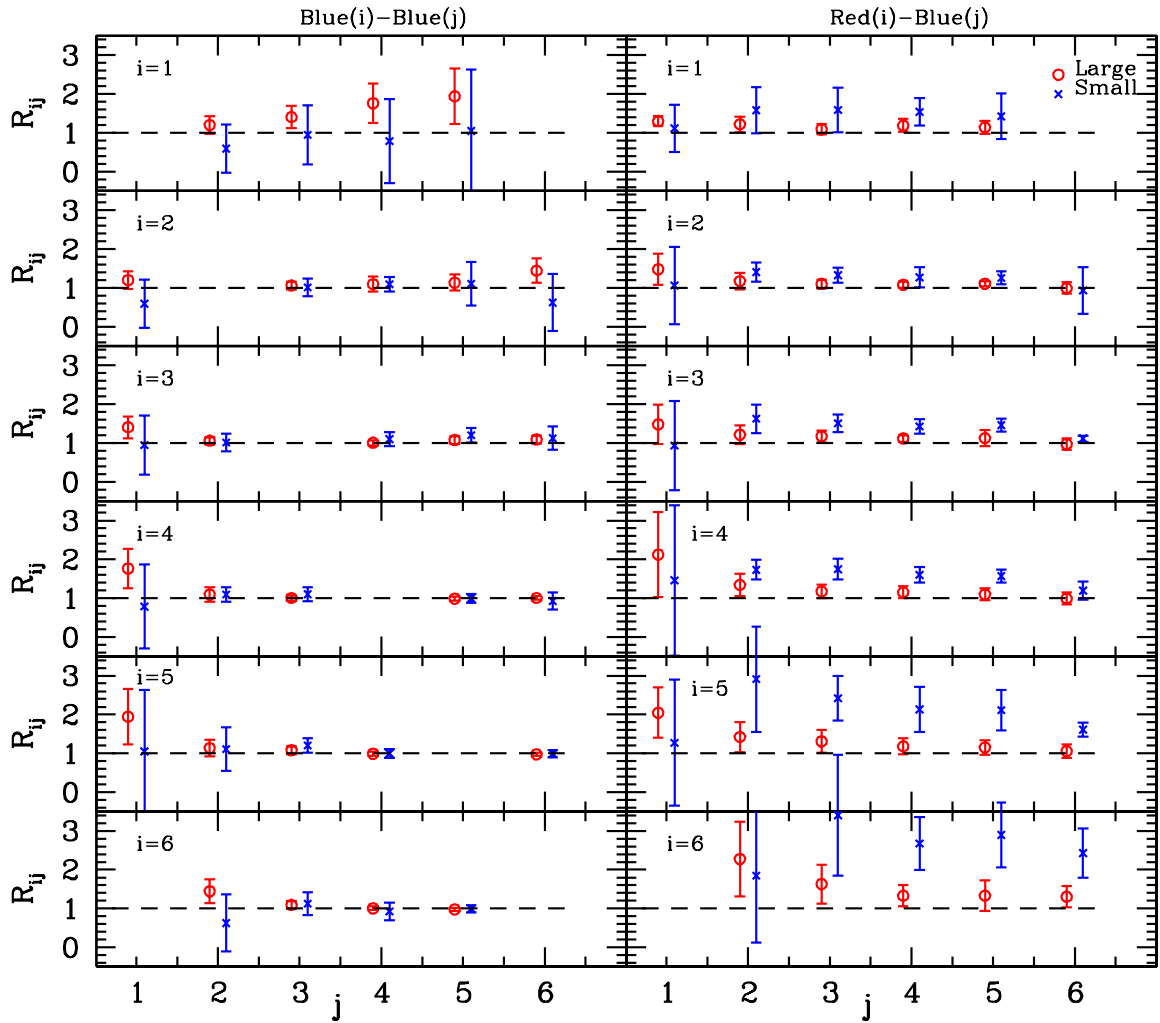}       \caption{The       same      as       Fig.\,
    \ref{fig:stochasticity1},  but  here results  are  shown for  blue
    versus blue  galaxies (left panels)  and red versus  blue galaxies
    (right panels).}
\label{fig:stochasticity2}
\end{figure*}

\begin{deluxetable*}{lcccccc}
  \tablecaption{The average value of ${\cal R}_{ij}$ over the large
  scales
    ($0.98\mpch \leq r_p \leq 9.6\mpch$)
\label{RijL}} \tablewidth{0pt}
  \tablehead{ Sample & ${\rm F1}$ & ${\rm F2}$ & ${\rm F3}$ & ${\rm F4}$ &
  ${\rm F5}$ & ${\rm F6}$ }

\startdata
${\rm F1}$ & &$ 1.00 \pm  0.04$&$ 1.05 \pm 0.06$&$ 1.05 \pm  0.11$&$ 1.14 \pm  0.08$& \\

${\rm F2}$ &$ 1.00 \pm  0.04$& &$ 1.13 \pm  0.06$&$ 1.02 \pm 0.03$&$ 1.12 \pm  0.02$&$ 0.91 \pm  0.14$\\
${\rm F3}$ &$ 1.05 \pm 0.06$&$ 1.13 \pm 0.06$& &$ 1.02 \pm 0.02$&$ 1.20 \pm  0.06$&$ 0.93 \pm  0.08$\\
${\rm F4}$ &$ 1.05 \pm 0.11$&$ 1.02 \pm 0.03$&$ 1.02 \pm 0.02$& &$ 1.30 \pm 0.13$&$ 0.93 \pm  0.05$\\
${\rm F5}$ &$ 1.14 \pm  0.08$&$ 1.12 \pm 0.02$&$ 1.20 \pm 0.06$&$ 1.30 \pm 0.13$& &$ 0.94 \pm 0.02$\\
${\rm F6}$ & &$ 0.91 \pm  0.14$&$ 0.93 \pm 0.08$&$ 0.93 \pm 0.05$&$ 0.94 \pm 0.02$& \\
\tableline \tableline
  & ${\rm F1_{red}}$ & ${\rm F2_{red}}$ & ${\rm F3_{red}}$ & ${\rm F4_{red}}$
  & ${\rm F5_{red}}$ & ${\rm F6_{red}}$  \\
 \tableline
${\rm F1_{red}}$ & &$ 0.98 \pm  0.05$&$ 0.98 \pm  0.05$&$ 1.05 \pm 0.19$&$1.09 \pm  0.12$& \\
${\rm F2_{red}}$ &$ 0.98 \pm 0.05$& &$ 1.02 \pm  0.05$&$1.04 \pm 0.06$&$ 1.06 \pm 0.05$&$ 1.17 \pm 0.27$\\
${\rm F3_{red}}$ &$ 0.98 \pm 0.05$&$ 1.02 \pm 0.05$& &$0.99 \pm  0.03$&$ 0.99 \pm  0.07$&$ 0.98 \pm  0.13$\\
${\rm F4_{red}}$ &$ 1.05 \pm  0.19$&$ 1.04 \pm  0.06$&$ 0.99 \pm  0.03$& &$ 0.97 \pm 0.04$&$ 1.03 \pm 0.08$\\
${\rm F5_{red}}$ &$ 1.09 \pm 0.12$&$ 1.06 \pm 0.05$&$ 0.99 \pm 0.07$&$0.97 \pm 0.04$& &$ 0.94 \pm 0.03$\\
${\rm F6_{red}}$ & &$ 1.17 \pm 0.27$&$ 0.98 \pm 0.13$&$1.03 \pm 0.08$&$ 0.94 \pm 0.03$& \\
\tableline \tableline
  & ${\rm F1_{blue}}$ & ${\rm F2_{blue}}$ & ${\rm F3_{blue}}$ & ${\rm
    F4_{blue}}$ & ${\rm F5_{blue}}$ & ${\rm F6_{blue}}$  \\
 \tableline
${\rm F1_{blue}}$ & &$ 1.20 \pm  0.23$&$ 1.41 \pm  0.28$&$ 1.76 \pm 0.51$&$ 1.94 \pm  0.71$& \\
${\rm F2_{blue}}$ &$ 1.20 \pm  0.23$& &$ 1.06 \pm  0.09$&$1.10 \pm 0.19$&$ 1.14 \pm  0.21$&$ 1.45 \pm  0.31$\\
${\rm F3_{blue}}$ &$ 1.41 \pm 0.28$&$ 1.06 \pm  0.09$& &$1.01 \pm 0.07$&$ 1.08 \pm  0.10$&$ 1.10 \pm  0.10$\\
${\rm F4_{blue}}$ &$ 1.76 \pm 0.51$&$ 1.10 \pm  0.19$&$ 1.01 \pm  0.07$& &$ 0.99 \pm  0.05$&$ 1.01 \pm  0.05$\\
${\rm F5_{blue}}$ &$ 1.94 \pm 0.71$&$ 1.14 \pm  0.21$&$ 1.08 \pm  0.10$&$0.99 \pm 0.05$& &$ 0.98 \pm 0.04$\\
${\rm F6_{blue}}$ & &$ 1.45 \pm  0.31$&$ 1.10 \pm 0.10$&$1.01 \pm 0.05$&$ 0.98 \pm  0.04$& \\
\tableline \tableline
  & ${\rm F1_{blue}}$ & ${\rm F2_{blue}}$ & ${\rm F3_{blue}}$ & ${\rm
    F4_{blue}}$ & ${\rm F5_{blue}}$ & ${\rm F6_{blue}}$  \\
 \tableline
${\rm F1_{red}}$ &$ 1.31 \pm  0.12$&$ 1.22 \pm  0.20$&$ 1.10 \pm  0.13$&$1.19 \pm  0.16$&$ 1.14 \pm  0.17$& \\
${\rm F2_{red}}$ &$ 1.49 \pm  0.40$&$ 1.18 \pm  0.21$&$ 1.10 \pm  0.11$&$1.08 \pm 0.08$&$ 1.11 \pm  0.06$&$ 1.00 \pm  0.15$\cr
${\rm F3_{red}}$ &$ 1.48 \pm 0.50$&$ 1.21 \pm  0.24$&$ 1.18 \pm  0.15$&$1.12 \pm 0.08$&$ 1.13 \pm  0.21$&$ 0.98 \pm  0.14$\cr
${\rm F4_{red}}$ &$ 2.12 \pm 1.10$&$ 1.35 \pm  0.28$&$ 1.18 \pm  0.17$&$1.16 \pm 0.15$&$ 1.10 \pm  0.15$&$ 1.00 \pm  0.16$\cr
${\rm F5_{red}}$ &$ 2.05 \pm 0.65$&$ 1.42 \pm  0.39$&$ 1.31 \pm  0.29$&$1.18 \pm 0.20$&$ 1.15 \pm  0.19$&$ 1.06 \pm 0.17$\cr
${\rm F6_{red}}$ & &$ 2.27 \pm  0.96$&$ 1.63 \pm 0.50$&$1.33 \pm 0.27$&$ 1.33 \pm  0.40$&$ 1.31 \pm 0.28$\cr
\enddata
\end{deluxetable*}

\begin{deluxetable*}{lcccccc}
  \tablecaption{The average value of ${\cal R}_{ij}$ over the small
  scales
    ($0.16\mpch \leq r_p \leq 0.98\mpch$)
\label{RijS}} \tablewidth{0pt}
  \tablehead{ Sample & ${\rm F1}$ & ${\rm F2}$ & ${\rm F3}$ & ${\rm F4}$ & ${\rm F5}$ & ${\rm F6}$ }
\startdata
${\rm F1}$ & &$ 0.92 \pm  0.10$&$ 0.92 \pm  0.08$&$ 0.80 \pm  0.34$&$ 0.81 \pm  0.44$& \\
${\rm F2}$ &$ 0.92 \pm 0.10$& &$ 1.07 \pm  0.11$&$ 1.00 \pm  0.07$&$ 1.23 \pm  0.23$&$ 1.03 \pm  0.53$\\
${\rm F3}$ &$ 0.92 \pm 0.08$&$ 1.07 \pm 0.11$& &$ 1.05 \pm 0.06$&$ 1.13 \pm  0.12$&$ 0.84 \pm  0.06$\\
${\rm F4}$ &$ 0.80 \pm 0.34$&$ 1.00 \pm 0.07$&$ 1.05 \pm 0.06$& &$ 1.12 \pm 0.07$&$ 0.90 \pm  0.11$\\
${\rm F5}$ &$ 0.81 \pm  0.44$&$ 1.23 \pm 0.23$&$ 1.13 \pm 0.12$&$ 1.12 \pm 0.07$& &$ 0.95 \pm 0.04$\\
${\rm F6}$ & &$ 1.03 \pm  0.53$&$ 0.84 \pm 0.06$&$ 0.90 \pm 0.11$&$ 0.95 \pm 0.04$& \\
\tableline \tableline
  & ${\rm F1_{red}}$ & ${\rm F2_{red}}$ & ${\rm F3_{red}}$ & ${\rm F4_{red}}$ & ${\rm F5_{red}}$ & ${\rm F6_{red}}$  \\
 \tableline
${\rm F1_{red}}$ & &$ 0.84 \pm  0.11$&$ 0.78 \pm  0.16$&$0.73 \pm  0.20$&$ 1.09 \pm  0.38$& \\
${\rm F2_{red}}$ &$ 0.84 \pm  0.11$& &$ 1.03 \pm  0.13$&$1.06 \pm 0.08$&$ 1.20 \pm  0.15$&$ 1.29 \pm  0.81$\\
${\rm F3_{red}}$ &$ 0.78 \pm 0.16$&$ 1.03 \pm  0.13$& &$1.06 \pm 0.09$&$ 1.08 \pm  0.14$&$ 1.06 \pm  0.32$\\
${\rm F4_{red}}$ &$ 0.73 \pm 0.20$&$ 1.06 \pm  0.08$&$ 1.06 \pm  0.09$& &$ 1.06 \pm  0.09$&$ 1.02 \pm  0.08$\\
${\rm F5_{red}}$ &$ 1.09 \pm 0.38$&$ 1.20 \pm  0.15$&$ 1.08 \pm  0.14$&$1.06 \pm 0.09$& &$ 1.01 \pm 0.08$\\
${\rm F6_{red}}$ & &$ 1.29 \pm  0.81$&$ 1.06 \pm 0.32$&$1.02 \pm 0.08$&$ 1.01 \pm  0.08$& \\
\tableline \tableline
  & ${\rm F1_{blue}}$ & ${\rm F2_{blue}}$ & ${\rm F3_{blue}}$ & ${\rm F4_{blue}}$ & ${\rm F5_{blue}}$ & ${\rm F6_{blue}}$  \\
 \tableline
${\rm F1_{blue}}$ & &$ 0.60 \pm  0.62$&$ 0.95 \pm  0.76$&$
0.79 \pm  1.08$&$ 1.05 \pm  1.58$& \\
${\rm F2_{blue}}$ &$ 0.60 \pm  0.62$& &$ 1.02 \pm  0.22$&$1.10 \pm 0.19$&$ 1.11 \pm  0.56$&$ 0.63 \pm  0.73$\\
${\rm F3_{blue}}$ &$ 0.95 \pm 0.76$&$ 1.02 \pm  0.22$& &$1.11 \pm 0.18$&$ 1.21 \pm  0.18$&$ 1.13 \pm  0.30$\\
${\rm F4_{blue}}$ &$ 0.79 \pm 1.08$&$ 1.10 \pm  0.19$&$ 1.11 \pm  0.18$& &$ 1.00 \pm  0.11$&$ 0.93 \pm  0.22$\\
${\rm F5_{blue}}$ &$ 1.05 \pm 1.58$&$ 1.11 \pm  0.56$&$ 1.21 \pm  0.18$&$1.00 \pm 0.11$& &$ 1.00 \pm 0.09$\\
${\rm F6_{blue}}$ & &$ 0.63 \pm  0.73$&$ 1.13 \pm 0.30$&$0.93 \pm 0.22$&$ 1.00 \pm  0.09$& \\
\tableline \tableline
  & ${\rm F1_{blue}}$ & ${\rm F2_{blue}}$ & ${\rm F3_{blue}}$ & ${\rm F4_{blue}}$ & ${\rm F5_{blue}}$ & ${\rm F6_{blue}}$  \\
 \tableline
${\rm F1_{red}}$ &$ 1.12 \pm  0.61$&$ 1.58 \pm  0.60$&$ 1.59 \pm  0.57$&$
1.54 \pm  0.35$&$ 1.43 \pm  0.59$& \\
${\rm F2_{red}}$ &$ 1.07 \pm  0.99$&$ 1.41 \pm  0.25$&$ 1.33 \pm  0.19$&$
1.28 \pm 0.26$&$ 1.26 \pm  0.17$&$ 0.94 \pm  0.60$\\
${\rm F3_{red}}$ &$ 0.94 \pm 1.15$&$ 1.63 \pm  0.37$&$ 1.51 \pm  0.23$&$
1.43 \pm 0.19$&$ 1.47 \pm  0.17$&$ 1.12 \pm  0.08$\cr
${\rm F4_{red}}$ &$ 1.46 \pm 1.94$&$ 1.74 \pm  0.25$&$ 1.75 \pm  0.27$&$
1.60 \pm 0.20$&$ 1.57 \pm  0.16$&$ 1.20 \pm  0.24$\cr
${\rm F5_{red}}$ &$ 1.28 \pm 1.62$&$ 2.91 \pm  1.35$&$ 2.42 \pm  0.57$&$
2.13 \pm 0.58$&$ 2.11 \pm  0.52$&$ 1.62 \pm 0.18$\cr
${\rm F6_{red}}$ & &$ 1.85 \pm  1.72$&$ 3.40 \pm 1.55$&$
2.68 \pm 0.68$&$ 2.90 \pm  0.83$&$ 2.43 \pm 0.63$\cr
\enddata
\end{deluxetable*}

In   the  last   two  sections   we  have   analyzed  the   auto-  and
cross-correlation functions for galaxies of different luminosities and
colors. We  have seen that  on scales $r_p\ga 1\mpch$  all correlation
functions  are  roughly  parallel,  but on  smaller  scales  different
galaxies  may  have different  behaviors  (see Fig.\,\ref{fig4}).   In
order to investigate the scale-dependence of the correlation functions
in  more detail,  we plot  in Fig.\,\ref{wrp}  the ratios  between the
various projected correlation functions, $W(r_p)$, and a normalization
function,  $W_0(r_p)$, which  is the  power-law fit  to  the projected
correlation  function  of  the   sample  $F^*$  on  large  scales  and
corresponds  to $r_0=5.49\mpch$ and  $\alpha=1.76$.  Three  top panels
show results for red galaxies;  six middle panels show results between
red and blue  galaxies; and three bottom panels  show results for blue
galaxies.  Based on the figure, we can draw the following conclusions:
\begin{itemize}
\item On  scales larger than  $1\mpch$, all the  correlation functions
  are roughly parallel, a fact we have mentioned earlier.
\item  The  cross-correlation  functions  between  the  brightest  red
  galaxies and  other red galaxies have enhanced  clustering on scales
  smaller than $\sim 1\mpch$, and the enhancement is the strongest for
  the cross-correlation between the brightest and faintest samples.
\item The  cross-correlation functions among faint  blue galaxies, and
  between faint-blue  and faint-red galaxies, appear  to be suppressed
  slightly on scales $\la  1\mpch$ relative to the extrapolations from
  larger scales.
\item The  brightest blue galaxies show enhanced  correlation with red
  galaxies of different luminosities on scales $\la 1\mpch$.
\end{itemize}
Some  of these  results are  expected, as  it is  well known  that red
galaxies reside  preferentially in rich systems, such  as clusters and
groups,  while faint  blue  galaxies are  distributed  in the  `field'
(i.e.,  in dark  matter haloes  of significantly  lower  mass).  Note,
however,  that the cross-correlation  functions between  the brightest
red   galaxies   and  faint   red   galaxies   are   lower  than   the
auto-correlation function of the  brightest galaxies on scales $r_p\ga
1\mpch$,  suggesting that not  all faint  red galaxies  are associated
with  the brightest  red galaxies.   There  is no  enhancement in  the
cross-correlation functions  on small scales between  red galaxies and
all except the brightest blue galaxies. This again can be explained by
the fact that blue galaxies with low and intermediate luminosities are
distributed in the field.

One surprising result is that the brightest blue galaxies are strongly
correlated with  red galaxies  on small scales,  and in many  ways the
clustering properties  of these  galaxies are similar  to that  of red
galaxies with  intermediate luminosities. In order  to understand this
result  in  more detail,  we  have  examined  the properties  of  this
population  of  galaxies  more  closely.  Using  the  group  catalogue
constructed by Yang et al.  (2007, in preparation), we find that $\sim
15\%$ ($\sim  80\%$) of  the brightest blue  galaxies are  the central
galaxies  of groups  with masses  $\ga 10^{13.5}h^{-1}  {\rm M}_\odot$
($\ga 10^{13}h^{-1}  {\rm M}_\odot$),  compared to $\sim  30\%$ ($\sim
85\%$)  for  the red  galaxies  in the  same  luminosity  bin.  It  is
therefore possible that a fraction  of the brightest blue galaxies are
in  fact `early-type' galaxies  and their  relatively blue  colors are
probably due  to rejuvenation of  star formation or AGN  activities in
their  centers.  Clearly,  further analysis  is required  in  order to
understand  the properties  of  this population  of  galaxies in  more
detail.

Next  we  discuss   how  our  results  may  put   constraints  on  the
stochasticity/non-linearity in the bias relation. Consider the density
field (smoothed on some scale) traced by a population $i$ of galaxies,
$\delta_i$.  In general we may write  it in terms of the density field
of another population, $\delta_j$, as
\begin{equation}
\delta_i = B_{ij}\delta_j + f_i(\delta_j)+ \epsilon_{ij}\,,
\end{equation}
where the first term on  the right-hand side describes the linear part
of the correlation between the two populations, the second term is the
non-linear part,  and $\epsilon_{ij}$ is  the stochastic, uncorrelated
part: $\langle\epsilon_{ij}\delta_j\rangle=0$.  It then follows that
\begin{equation}
\langle \delta_i\delta_j\rangle
= B_{ij}\langle \delta_j^2\rangle
 +\langle\delta_j f_i (\delta_j)\rangle\,,
\end{equation}
and
\begin{equation}
\langle \delta_i^2\rangle
= B_{ij}^2 \langle \delta_j^2\rangle
+\langle\epsilon_{ij}^2\rangle
+\langle f_i^2(\delta_j)\rangle
+2B_{ij}\langle\delta_j f_i(\delta_j)\rangle \,.
\end{equation}
Thus, for any two populations we can define two relative bias factors,
one based  on the cross  correlation with another population,  and one
based on the auto-correlation functions of the two populations:
\begin{equation}
B^{\rm (c)}_{ij}
\equiv {\langle\delta_i\delta_j\rangle
\over\langle\delta_j^2\rangle}\,;
\end{equation}
\begin{equation}
B^{\rm (a)}_{ij} \equiv \left({\langle\delta_i^2\rangle
\over\langle\delta_j^2\rangle}\right)^{1/2}\,.
\end{equation}
The square of the ratio between these two bias factors is
\begin{eqnarray}
\label{ratio:bias}
{\cal R}_{ij} &\equiv& \left({B_{ij}^{\rm (a)}\over
B_{ij}^{\rm (c)}}\right)^2
={\langle\delta_i^2\rangle\langle\delta_j^2\rangle
 \over \langle\delta_i\delta_j\rangle^2}\nonumber\\
&=&
1+{\langle \epsilon_{ij}^2\rangle + \zeta_{ij}
\over \langle\delta_i\delta_j\rangle^2
/\langle\delta_j^2\rangle}\,,
\end{eqnarray}
where
\begin{equation}
\zeta_{ij}\equiv
\langle f_i^2(\delta_j)\rangle
-\langle\delta_j f_i(\delta_j)\rangle^2/\langle\delta_j^2\rangle
\end{equation}
describes the contribution due to  the non-linear part of the relative
bias  relation.  Note  that  ${\cal  R}$  is  related  to  the  linear
correlation coefficient,  $r$, defined in Dekel \&  Lahav (1999; their
eq.[13]) as  ${\cal R}  = r^{-2}$.  For  two populations, $i$  and $j$
that  are  entirely  uncorrelated  one has  that  $R_{ij}  \rightarrow
\infty$, while  $R_{ij}=1$ when $i$ and $j$  are perfectly correlated.
Both the  stochastic component, $\langle  \epsilon_{ij}^2\rangle$, and
the non-linear  component, $\zeta_{ij}$, can cause  ${\cal R}_{ij}$ to
deviate  from  unity, and  in  general  it  is difficult  to  separate
stochasticity  from  non-linearity.   On  large scales,  however,  the
non-linear component  in the  bias relation may  be neglected,  and we
have that
\begin{equation}
{\cal R}_{ij} = 1+
{\langle\epsilon_{ij}^2\rangle \over B_{ij}^2
\langle\delta_j^2\rangle}\,.
\end{equation}

To  examine  the  importance  of  stochasticity/non-linearity  in  the
relative bias relations, we  calculate ${\cal R}_{ij}$ for galaxies of
different    luminosities    and    colors.    Based    on    equation
(\ref{ratio:bias}), we estimate ${\cal R}_{ij}$ using
\begin{equation}
{\cal R}_{ij} = {W_{ii}W_{jj} \over W_{ij}^2}\,,
\end{equation}
where  $W_{ii}$ and  $W_{jj}$  are the  auto-correlation functions  of
subsamples `$i$'  and `$j$', respectively, and $W_{ij}$  is the cross-
correlation function  between the two subsamples. Note  that `$i$' and
`$j$'  denote not  only luminosity  samples  but also  color samples.
Unlike  what we  did above,  here we  estimate $W_{ii}$,  $W_{jj}$ and
$W_{ij}$ all in the overlapping region of the two samples in question,
so that volume effects are minimized.

In Fig\,\ref{fig_Rij} we  show ${\cal R}_{ij}$ as a  function of $r_p$
for  various  combinations  of  color-  and  luminosity-samples.   The
results  are quite noisy  in some  cases, again  because of  the small
overlap  of the samples  involved.  As  one can  see, for  cases where
${\cal R}_{ij}$ is  well determined, it is close to  1 for all-all and
red-red pairs of samples. For blue-blue pairs, ${\cal R}_{ij}$ is also
close to 1 except the  cases involving the brightest galaxies where at
large scales ${\cal  R}_{ij}$ can be significantly larger  than 1.  As
we have  discussed above, the brightest blue  galaxies have clustering
properties  different from that  of fainter  blue galaxies,  because a
significant fraction  of them are,  like bright red  galaxies, central
galaxies of relatively massive  groups.  For red versus blue galaxies,
${\cal R}_{ij}$ is  significantly larger than 1 on  small scales.  The
signal  is  also more  prominent  between  faint  red and  faint  blue
galaxies and can extend to a few $\mpch$.

To  show  this more  clearly,  we plot  the  average  value of  ${\cal
R}_{ij}$  over  two  different  scales.   The  results  are  shown  in
Figs.~\ref{fig:stochasticity1} --  \ref{fig:stochasticity2} and listed
in Tables  ~\ref{RijL} --  \ref{RijS} for all  versus all,  red versus
red, blue versus blue, and  red versus blue, respectively.  In all the
figures,  circles show  results based  on clustering  on  large scales
($0.98\mpch \leq r_p \leq 9.6\mpch$), while crosses show results based
on clustering  on small scales ($0.16\mpch \leq  r_p \leq 0.98\mpch$).
As one can see, on large scales ($\ga 1\mpch$) ${\cal R}_{ij}$ is very
close  to  1 for  red-red,  blue-blue  (with  the exception  of  cases
involving the brightest  blue galaxies), and all-all pairs,  but has a
value  between $1.4$ to  $2.2$ for  pairs between  faint red  and blue
galaxies.  On scales $\la 1\mpch$, the  ratio is again very close to 1
for all-all, red-red and  blue-blue pairs, but is significantly larger
than 1 for  red versus blue galaxies  and is as large as  $\sim 3$ for
faint-red versus faint-blue galaxies.

These  results  suggest  that  red  (or blue)  galaxies  of  different
luminosities  trace  each  other  almost  linearly and  in  an  almost
deterministic  way on  both small  and large  scales, but  there  is a
significant   stochastic/non-linear  component  in   the  relationship
between  red  and blue  galaxies  of  different  luminosities, and  in
particular  on small  scales. The  large  value of  $R_{ij}$ on  small
scales  between  red and  blue  galaxies  (except  the brightest  blue
galaxies) is  almost certainly due to  the fact that  blue galaxies of
low  and intermediate  luminosities  avoid dense  groups and  clusters
while  red galaxies reside  preferentially in  such places.   The fact
that  such stochastic/non-linear  component extends  to a  few $\mpch$
(see Fig.~\ref{fig_Rij}) suggests that  the clustering of clusters and
groups  on  large  scales   may  produce  segregation  of  the  galaxy
population on large scales.

\section{Summary}
\label{sec:summary}

In this  paper, we use galaxy  samples constructed from  the SDSS
Data Release  4  (DR4) to  study  the  cross-correlation functions
between galaxies  with  different   luminosities  and  colors.
Galaxies  are separated  at  $g-r=-0.7-0.032(M_{r,0.1}+16.5)$   into
red  and  blue samples  according to their  colors, and  each  of
the color samples  is further divided into 6  subsamples according
to luminosity. We measure the projected  two-point correlation
function for  each subsample, and the  projected cross-correlation
for  each pair  of subsamples.  Each correlation  function is
fitted with power laws  over  two different ranges  of  separations:
$0.98\mpch \leq  r_p  \leq  9.6\mpch$  and $0.16\mpch \leq  r_p \leq
0.98\mpch$, and  each of the  power laws are characterized by a
correlation length, $r_0$, and a logarithmic slope, $\alpha$. Our
main results can be summarized as follows:
\begin{enumerate}
\item At projected separations $r_p \ga 1\mpch$, all correlation functions are
  roughly parallel to each other (but see exceptions for the brightest
  galaxies with $-22.0\ge M_{r,0.1} \ge -23.0$ in Zehavi et al. 2005; Li et
  al. 2006), but on smaller scales different populations of galaxies have
  different behaviors.
\item  The auto- and  cross-correlation functions  of red  galaxies on
  $r_p  \la 1\mpch$ are  significantly enhanced  relative to  those on
  larger scales. The effect is the strongest between the brightest red
  galaxies  and other red  galaxies.  Such  enhancement is  absent for
  blue  galaxies and  in the  cross-correlation between  red  and blue
  galaxies.
\item For  blue galaxies the luminosity-dependence  of the correlation
  amplitude  on large  scales  is strong  over  the entire  luminosity
  range, while for  red galaxies the dependence is  weaker and becomes
  insignificant for  luminosities below $L^*$. There  is an indication
  that the luminosity-dependence may inverse at the faint end.
\item For a given luminosity, red galaxies are more strongly clustered
  than  blue  galaxies  on  both  large  and  small  scales,  and  the
  difference is larger for fainter luminosities.
\item  One large  scale, the  bias factors  of a  given  population of
  galaxies obtained from its cross correlations with other populations
  of  galaxies  are similar,  and  comparable  to  that based  on  the
  auto-correlation function.
\item On  both large  and small scales,  the ratio ${\cal  R}_{ij}$ is
  very close to  1 for red (or blue except  the brightest) galaxies of
  different luminosities,  suggesting that the  bias relations between
  red (or blue) galaxies of different luminosities are close to linear
  and deterministic.
\item On scales $\la 1\mpch$,  $R_{ij}$ is significantly larger than 1
  between red and blue galaxies and as large as $\sim 2$ between faint
  red  and   faint  blue  galaxies,  suggesting   that  a  significant
  stochastic/non-linear component exists in the bias relations between
  blue and red galaxies on small scales.
\item  The  clustering  properties  of  the  brightest  blue  galaxies
  resemble  more that  of red  galaxies of  intermediate luminosities
  than that  of fainter blue  galaxies, indicating that  a significant
  fraction of them are located in relative rich systems.
\end{enumerate}

The results obtained in this paper provide important information about the
relationships between the spatial distributions of galaxies of different
properties in space. Part of our findings are consistent with previous probes
(e.g. the luminosity and color dependences of SDSS galaxies based on the {\it
  auto} correlation functions by Zehavi et al. 2005; Li et al. 2006). Most of
our findings will put new constraints on the galaxy formation models.  To
explore the implications of our findings for galaxy formation in the cosmic
density field, the results obtained here should be combined with
semi-analytical and/or halo occupation models to constrain how galaxies
populate dark matter halos, and how galaxies are distributed in individual
dark matter halos. We will come back to this in the future.

During the final stages of this project a paper appeared by Swanson et
al.   (2007), who  carried out  a similar  investigation to  study the
relative  biases for galaxies  of different  luminosities and  colors.
Although their analysis is based on counts-in-cells, as opposed to the
correlation  functions  used  here,  their results  are  qualitatively
similar  to  ours.  A  detailed  comparison,  however, is  complicated
because  of the  differences  in the  statistical  methods and  sample
selections.


\acknowledgments We thank Cheng Li for the help in understanding the SDSS
data. XY is supported by the {\it One Hundred Talents} project and the
Knowledge Innovation Program (Grant No.  KJCX2-YW-T05) of the Chinese Academy
of Sciences and grants from NSFC (Nos.10533030, 10673023).  HJM would like to
acknowledge the support of NSF AST-0607535, NASA AISR-126270 and NSF
IIS-0611948.

Funding for  the SDSS and SDSS-II has  been provided by the  Alfred P.
Sloan Foundation, the Participating Institutions, the National Science
Foundation, the  U.S.  Department of Energy,  the National Aeronautics
and Space Administration, the  Japanese Monbukagakusho, the Max Planck
Society, and  the Higher Education  Funding Council for  England.  The
SDSS Web  Site is  http://www.sdss.org/.  The SDSS  is managed  by the
Astrophysical Research Consortium  for the Participating Institutions.
The  Participating Institutions  are  the American  Museum of  Natural
History,  Astrophysical   Institute  Potsdam,  University   of  Basel,
Cambridge University,  Case Western Reserve  University, University of
Chicago,  Drexel  University,  Fermilab,  the Institute  for  Advanced
Study, the  Japan Participation  Group, Johns Hopkins  University, the
Joint  Institute for  Nuclear  Astrophysics, the  Kavli Institute  for
Particle Astrophysics  and Cosmology, the Korean  Scientist Group, the
Chinese Academy of Sciences  (LAMOST), Los Alamos National Laboratory,
the     Max-Planck-Institute     for     Astronomy     (MPIA),     the
Max-Planck-Institute   for  Astrophysics   (MPA),  New   Mexico  State
University,   Ohio  State   University,   University  of   Pittsburgh,
University  of  Portsmouth, Princeton  University,  the United  States
Naval Observatory, and the University of Washington.


\begin{thebibliography}{}

\bibitem[\protect\citeauthoryear{{Adelman-McCarthy}, {Ag{\"u}eros},
  {Allam}, {Anderson}, {Anderson}, {Annis}, {Bahcall}, {Baldry} \& {et
  al.,}}{{Adelman-McCarthy} et~al.}{2006}]{Adelman-McCarthy-06}
  {Adelman-McCarthy} J.~K.,  {Ag{\"u}eros} M.~A.,  {Allam} S.~S.,
  {Anderson} K.~S.~J.,  {Anderson} S.~F.,  {Annis} J.,  {Bahcall}
  N.~A.,  {Baldry} I.~K., {et al.,} 2006, \apjs, 162, 38

\bibitem[]{Baldry04}
Baldry I.K., Glazebrook K., Brinkmann J., Ivezi\'{c} \v{Z}., Lupton
R.H., Nichol R.C., Szalay A.S., 2004, \apj, 600, 681

\bibitem[]{Bell04}
Bell E.F., et al., 2004, \apj, 608, 752

\bibitem[Benoist et al.(1996)]{benoist96}
Benoist, C., Maurogordato, S., da Costa, L. N., Cappi, A., \&
Schaeffer, R. 1996, \apj, 472, 452

\bibitem[\protect\citeauthoryear{Berlind \& Weinberg}{2002}]{2002ApJ...575..587B}
Berlind A. A., Weinberg D.~H., 2002, ApJ, 575, 587

\bibitem[\protect\citeauthoryear{{Blanton}, {Brinkmann}, {Csabai},
  {Doi}, {Eisenstein}, {Fukugita}, {Gunn}, {Hogg} \& {et
  al.,}}{{Blanton} et~al.}{2003a}]{Blanton-03-Kcorrection}
  {Blanton}  M.~R.,  {Brinkmann} J.,  {Csabai} I.,  {Doi} M.,  {Eisenstein} D.,
  {Fukugita} M.,  {Gunn} J.~E.,  {Hogg} D.~W.,    {et al.,} 2003a,
  \aj, 125, 2348

\bibitem[\protect\citeauthoryear{{Blanton}, {Hogg}, {Bahcall},
  {Brinkmann}, {Britton}, {Connolly}, {Csabai}, {Fukugita} \& {et
  al.,}}{{Blanton} et~al.}{2003b}]{Blanton-03-LF} {Blanton} M.~R.,
  {Hogg} D.~W.,  {Bahcall} N.~A.,  {Brinkmann} J.,  {Britton} M.,
  {Connolly} A.~J.,  {Csabai} I.,  {Fukugita} M.,    {et al.,} 2003b,
  \apj, 592, 819

\bibitem[]{Blanton03c}
Blanton M.~R.,  {et al.,} 2003c, \apj, 594, 186

\bibitem[\protect\citeauthoryear{Blanton etal }{2005}]{2005AJ....129.2562B}
Blanton M. R., et al., 2005, AJ, 129, 2562 

\bibitem[Brown, Webster \& Boyle(2000)]{brown00}
Brown, M.\ J.\ I., Webster, R. L., \& Boyle, B.\ J.\ 2000, \mnras, 317, 782

\bibitem[Boerner et al.(1989)]{1989A&A...221..191B}
B\"orner, G., Mo, H., \& Zhou, Y.\ 1989, A\&A, 221, 191

\bibitem[Budavari et al.(2003)]{budavari03}
Budavari, T., et al., 2003, \apj, 595, 59

\bibitem[\protect\citeauthoryear{Cooray \& Sheth}{2002}]{2002PhR...372....1C}
Cooray A., Sheth R., 2002, PhR, 372, 1

\bibitem[Peebles(1983)]{peebles83}
Davis, M., Peebles, P.J.E.\ 1983, ApJ 267,465

\bibitem[]{Dekel99}
Dekel A., Lahav O., 1999, \apj , 520, 24

\bibitem[Guzzo et al.(1997)]{guzzo97}
Guzzo, L., Strauss, M.\ A., Fisher, K.\ B., Giovanelli, R., \&
Haynes, M.\ P.\ 1997, \apj, 489, 37

\bibitem[Goto et al.(2003)]{goto03}
Goto, T., Yamauchi, C., Fujita, Y., Okamura, S., Sekiguchi, M.,
Smail, I., Bernardi, M., \& Gomez, P.\ L. 2003, \mnras 346, 601

\bibitem[Hamilton \& Tegmark(2004)]{ht04}
Hamilton, A.~J.~S.~\& Tegmark, M.\ 2004, \mnras, 349, 115

\bibitem[]{Hogg04}
Hogg D.W., et al., 2004, \apj, 601, L29

\bibitem[Jing et al.(1991)]{1991A&A...252..449J}
Jing, Y.~P., Mo, H.~J., \& Boerner, G.\ 1991, A\&A, 252, 449

\bibitem[Jing, Mo, \& B{\" o}rner(1998)]{jmb98}
Jing, Y. P., Mo, H. J., \& B{\" o}rner, G.\ 1998, \apj, 494, 1

\bibitem[Li et al.(2006)]{Li06}
Li, C., Kauffmann, G., Jing, Y.P., White, S. D. M., Boerner, G.,
Cheng, F.Z. 2006, \mnras, 368, 21

\bibitem[Loveday et al.(1995)]{loveday95}
Loveday, J., Maddox, S. J., Efstathiou, G., \& Peterson, B. A. 1995,
\apj, 442, 457

\bibitem[Madgwick et al.(2003)]{madgwick03}
Madgwick, D.~S. et al. 2003, \mnras, 344, 847

\bibitem[]{}
Mo H.J., White S.D.M., 1996, 282, 347

\bibitem[Norberg et al.(2001)]{norberg01}
Norberg, P., et al. 2001, \mnras, 328, 64

\bibitem[Norberg et al.(2002)]{norberg02}
Norberg, P., et al. 2002, \mnras, 332, 827

\bibitem[Park et al.(1994)]{park94}
Park, C., Vogeley, M. S., Geller, M. J., \& Huchra, J. P. 1994,
\apj, 431, 569

\bibitem[\protect\citeauthoryear{Peacock \& Smith}{2000}]{2000MNRAS.318.1144P}
Peacock J.~A., Smith R.~E., 2000, MNRAS, 318, 1144

\bibitem[Peebles(1980)]{peebles80}
Peebles, P.J.E.\ 1980, The Large-Scale Structure of the Universe,
Princeton University Press, Princeton

\bibitem[\protect\citeauthoryear{Seljak}{2000}]{2000MNRAS.318..203S}
Seljak U., 2000, MNRAS, 318, 203

\bibitem[]{Swanson07}
Swanson M.E.C., Tegmark M., Blanton M., Zehavi I., 2007, preprint (astro-ph/0702584)

\bibitem[\protect\citeauthoryear{van den Bosch, Mo, \& Yang}{2003}]{2003MNRAS.340.771B}
van den Bosch, F. C., Yang, X., Mo, H. J., 2003, MNRAS, 340, 771

\bibitem[Weinmann(2006)]{Weinmann2006}
Weinmann, S.M., van den Bosch, F. C., Yang, X.H., Mo, H. J., Croton,
D. J., Moore, B., 2006, \mnras, 366, 2

\bibitem[Willmer, da Costa \& Pellegrini(1998)]{willmer98}
Willmer, C. N. A., da Costa, L. N., \& Pellegrini, P. S. 1998, \aj,
115, 869

\bibitem[\protect\citeauthoryear{Yang, Mo, \& van den Bosch}{2003}]{2003MNRAS.339.1057Y}
Yang, X., Mo, H. J., van den Bosch, F. C., 2003, MNRAS, 339, 1057

\bibitem[\protect\citeauthoryear{Yang, Mo, \& van den Bosch}{2005}]{2005MNRAS}
Yang, X., Mo, H. J., van den Bosch, F. C., Weinmann, S.M., C., Li, Y.P.,
Jing, 2005, MNRAS, 362, 711

\bibitem[York etal (2000)]{york00}
York, D.~G., et al., 2000, \aj, 120, 1579

\bibitem[Zehavi etal (2002)]{zehavi02}
Zehavi, I., et al., 2002, \apj, 571, 172

\bibitem[\protect\citeauthoryear{Zehavi et al.}{2005}]{2005ApJ...630....1Z}
Zehavi I., et al., 2005, ApJ, 630, 1

\end{thebibliography}
\end{document}